\documentclass{nature_fig}
\usepackage{Preamble}
\title{Uniaxial-strain Control of Nematic Superconductivity in \sbs}

\author{
Ivan~Kostylev$^{1\ast}$, Shingo~Yonezawa$^{1\ast}$, Zhiwei~Wang$^{2,3}$,
Yoichi~Ando$^{2}$, Yoshiteru~Maeno$^1$}

\begin{document}

\maketitle

\vspace{-0.5cm} 

\begin{affiliations}
{\small
 \item Department of Physics, Graduate School of Science, Kyoto University, Kyoto 606-8502, Japan
 \item Institute of Physics II, University of Cologne, K\"{o}ln 50937, Germany
 \item Key Laboratory of Advanced Optoelectronic Quantum Architecture and
Measurement, Ministry of Education (MOE), School of Physics, Beijing
Institute of Technology, Beijing 100081, P. R. China
}
\end{affiliations}

%\vspace{-0.3em}

\noindent
$^{\ast}$e-mail: kostylev@scphys.kyoto-u.ac.jp, yonezawa@scphys.kyoto-u.ac.jp

\vspace{-1.5em}

 \noindent

\vspace{1em}

%\linenumbers

%FOR LETTER STYLE: This format begins with a title of, at most, 15 words, followed by an introductory paragraph (not abstract) of approximately 150 words, summarizing the background, rationale, main results (introduced by "Here we show" or some equivalent phrase) and implications of the study. This paragraph should be referenced, as in Nature style, and should be considered part of the main text, so that any subsequent introductory material avoids too much redundancy with the introductory paragraph.

\begin{abstract}
%1) Introduction: short general intro, detailed into, previous work that delineates the gap in our knowledge (2~3 sentences)
Nematic states are characterized by rotational symmetry breaking without translational ordering~\cite{nematicelectron1998,Fradkin2010,Fisher2012,Fernandes2014}. Recently, nematic superconductivity, in which the superconducting gap spontaneously lifts the rotational symmetry of the lattice, has been discovered~\cite{Fu2014, Matano2016.NaturePhys.12.852, Yonezawa2017.NaturePhys.13.123, Pan2016.SciRep.6.28632, Asaba2017.PhysRevX.7.011009, Yonezawa2019.CondensMatter.4.2}. However the pairing mechanism and the mechanism determining the nematic orientation remain unresolved. A first step is to demonstrate control of the nematicity, through application of an external symmetry-breaking field, to determine the sign and strength of coupling to the lattice. Here, we report for the first time control of the nematic orientation of the superconductivity of Sr$_x$Bi$_2$Se$_3$, through externally-applied uniaxial stress. The suppression of subdomains indicates that it is the $\Delta_{4y}$ state that is most favoured under compression along the basal Bi-Bi bonds. These results provide an inevitable step towards understanding the microscopic origin of the unique topological nematic superconductivity.

% Nematic states are characterized by rotational-symmetry breaking without translational ordering, accompanying high controllability of the order parameter~\cite{nematicelectron1998,Fradkin2010,Fisher2012,Fernandes2014}.
% Recently, nematic superconductivity, with unusual rotational-symmetry breaking in the superconducting gap amplitude, was discovered~\cite{Fu2014, Matano2016.NaturePhys.12.852, Yonezawa2017.NaturePhys.13.123, Pan2016.SciRep.6.28632, Asaba2017.PhysRevX.7.011009, Yonezawa2019.CondensMatter.4.2}. 
% However, fundamental issues such as the Cooper-pair glue or the mechanism determining the nematic director, remain unresolved. Finding strong coupling between the nematic superconductivity and an external symmetry breaking field, through experimental demonstration of high controllability, will provide important bases towards clarification of such issues.
% Here we show the first report on the control of nematic superconductivity in \sbs, via external uniaxial strain.
% By applying uniaxial strain in situ, we reversibly controlled the superconducting nematic domain structure.
% The suppression of subdomains indicates that the \Dy~state is most favored under compression. 
% This fact determines the coupling parameter between nematic superconductivity and uniaxial distortion, providing an inevitable step towards understanding the microscopic origin of the unique topological nematic superconductivity.% [149 words / 150 word limit]
\end{abstract}

%\clearpage %Comment out

%\startcoloring{Fig} %Comment out

%\section{[Introduction]} %Comment out

%\textbf{[Intro; nematic order in general]} 
%Nematic orders have been widely studied in condensed-matter physics. 
In nematic states of liquid crystals, bar-shaped molecules exhibit orientational ordering and forms thread-like topological defects of the order parameter. Because of the peculiar ``partial ordering'' property, the orientation of the molecules and hence the structure around defects are easily controlled by external stimuli, as widely utilized in liquid-crystal displays. Analogous phenomena in electronic systems, nematic electron liquids, where conduction electrons exhibit orientational ordering, have been revealed~\cite{nematicelectron1998,Fradkin2010}. Here, orientational properties are also highly controllable, and observations of such tunability have played fundamental roles to clarify driving mechanisms~\cite{Fisher2012,Fernandes2014}. 

%\textbf{[Intro: Nematic SC]}
A more exotic form of nematicity has been discovered in \textit{A}$_x$Bi$_2$Se$_3$ (\textit{A} = Cu, Sr, Nb)~\cite{Fu2014,Matano2016.NaturePhys.12.852,Yonezawa2017.NaturePhys.13.123,Pan2016.SciRep.6.28632,Asaba2017.PhysRevX.7.011009}: nematic superconductivity\cite{Fu2014}, in which the superconducting (SC) gap amplitude spontaneously lifts the rotational symmetry of the lattice. A consensus has been established that the gap has two-fold rotational symmetry, while the lattice has a three-fold rotational symmetry~\cite{Yonezawa2019.CondensMatter.4.2}. However for definitive demonstration it is essential to show control over the nematic orientation.

% A more exotic form of nematicity --- nematic superconductivity \cite{Fu2014} --- characterized by the unusual rotational-symmetry breaking in the superconducting (SC) gap amplitude, has been discovered in \abs~(\textit{A} = Cu, Sr, Nb)~\cite{Matano2016.NaturePhys.12.852,Yonezawa2017.NaturePhys.13.123,Pan2016.SciRep.6.28632,Asaba2017.PhysRevX.7.011009}. Although this phenomenon is rapidly gaining consensus in the community, fundamental issues remain unresolved~\cite{Yonezawa2019.CondensMatter.4.2}. Control of nematic superconductivity by an external symmetry-breaking field, e.g. uniaxial strain, leads to unambiguous demonstration of direct coupling between the nematicity and symmetry-breaking field, and thus provides important bases toward clarification of the issues. However, there are no experimental reports on such control of nematic superconductivity.

%\textbf{[Summary of this Letter]} 
In this Letter, we report the first control of nematic superconductivity in Sr$_x$Bi$_2$Se$_3$, through application of in situ tunable uniaxial stress along the $a$ axis (meaning a Bi-Bi bond direction). We reversibly controlled the nematic domain structure, allowing us to determine the sign of the coupling constant between the nematicity and lattice distortion.
%In this Letter, we report the first control of nematic superconductivity in \sbs, via the in-situ application of external uniaxial strain along the $a$ axis. We reversibly controlled the nematic SC domain structure, the primary topological defect, giving the experimental determination of the coupling-constant sign between the nematic superconductivity and uniaxial distortion.

\begin{figure}
\begin{center}
\includegraphics[width=14cm]{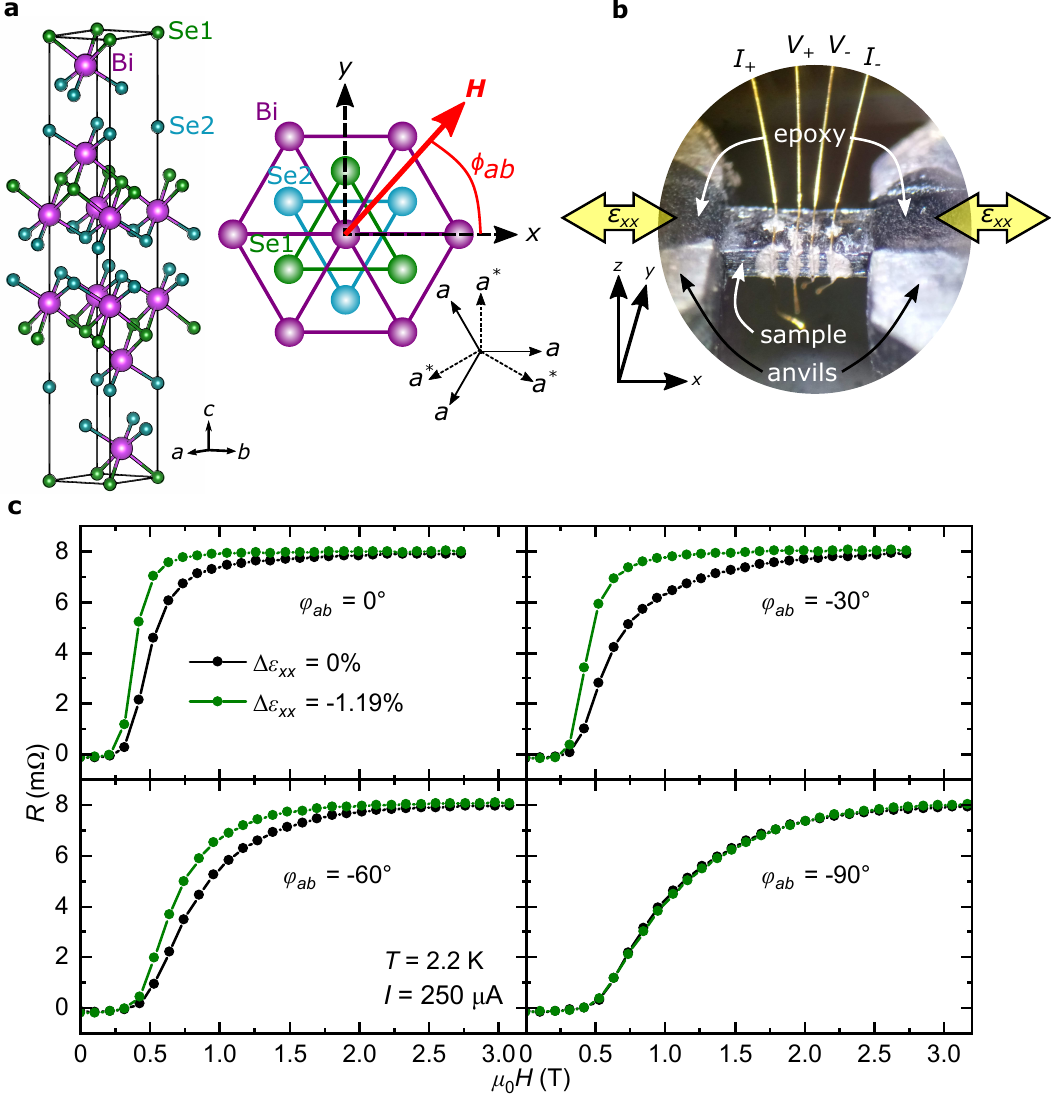}
\end{center}
\caption{
{\bf Uniaxial-strain control of nematic superconductivity in \sbs.}
{\bf a}, Crystal structure of the mother compound \bs. The right figure shows the definitions of the axes and the field angle \phiab\ with respect to the crystal structure in the $ab$ plane, with three equivalent \textit{a}- and \textit{a}*-axes.
{\bf b}, Photograph of the sample in the uniaxial strain cell with 4-wire terminal configuration. \textit{I} and \textit{V} labels next to the gold wires indicate the current and voltage leads, respectively. The large yellow arrows indicate the direction of the external strain, which was applied parallel to the \textit{x} axis.
{\bf c}, Magnetoresistance at specified field directions in the \textit{ab} plane (\phiab=0\deg, $-30$\deg, $-60$\deg, $-90$\deg), with and without $\Delta\varepsilon_{xx}$. The data were obtained at 2.2~K and with \SI{250}{\micro\ampere} applied current. A substantial change in the magnetoresistance curves under large $\Delta\varepsilon_{xx}$ (green curves) provides evidence for the in-situ uniaxial-strain control of the nematic superconductivity.
\label{fig1}
}
\end{figure}

%\textbf{[Intro: more explanation on doped \BS]} 
Our target materials family \abs~is derived from the topological insulator \BS~\cite{Xia2009.NaturePhys.5.398,Hsieh2009.Nature.460.1101}, which has a trigonal crystalline symmetry with three equivalent crystalline $a$ axes in the basal plane (Fig.~\ref{fig1}a)\cite{Lind2003}. Because the superconductivity induced by $A$ ion intercalation\cite{Hor2010.PhysRevLett.104.057001,Shruti2015.PhysRevB.92.020506,Qiu2015.arXiv:1512.03519.Full} occurs in its topologically non-trivial bands\cite{Wray2010.NaturePhys.6.855, Lahoud2013.PhysRevB.88.195107}, the resultant superconductivity can also be topologically non-trivial. Indeed, topological SC states have been proposed, among which a pair of SC states in the two-dimensional $E_u$ representation, \Dx~and \Dy, are nematic SC states~\cite{Fu2010.PhysRevLett.105.097001,Sasaki2011.PhysRevLett.107.217001,Fu2014,Sato2017.RepProgPhys.80.076501}. The SC gap amplitude of the \Dx~and \Dy~states are two-fold anisotropic and their maximum amplitude is located along the $a$ and $a^*$ axes, respectively (Fig.~\ref{fig1}a).
It has been shown that there is sample-to-sample variation in whether the nematicity aligns along a $a$ or $a^*$ axis~\cite{Du2017.SciChinaPhysMechAstron.60.037411}.
This fact suggests that \Dx\ and \Dy\ states are nearly degenerate states, such that the preferred state can be selected by a certain pre-existing symmetry breaking field such as possible structural distortion or \textit{A} ion distribution. Here, we probe whether applied uniaxial stress can overcome this pinning.

%\textbf{[Intro: pairing mechanism of doped \BS]} 
Another important issue is the pairing mechanism in \abs. This material is quite exceptional in the sense that it exhibits unconventional superconductivity without any proximity to magnetic or electric orderings and without strong electron-electron correlations\cite{Kriener2011.PhysRevLett.106.127004}. 
%There are theoretical proposals of nematic SC states mediated by phonons propagating along the [001] direction accompanying odd-parity fluctuations\cite{wan2014turning,PhysRevB.90.184512,PhysRevB.96.144504,Wang2019}. 
Uniaxial strain effects provides hints toward examination of various pairing scenarios.

%\section{[Results]} %Comment out

%\textbf{[Brief explanation of experimental setup]} 
In this work, we measured the magnetoresistance of single-crystalline \sbsn~samples (with the critical temperature \Tc~of 2.8 K; see Supplementary Note~\ref{sec:Supp_RvsT_dExx}) under uniaxial strain. The sample was affixed onto a custom-made uniaxial strain cell\cite{JAPKostylev2019}, a modified version of the recent invention~\cite{Hicks2014RSI}, mounted inside a vector magnet. The sample was cut along one of the \textit{a}~axes (Bi-Bi bond direction), which we define as the \textit{x}~axis (Figs.~\ref{fig1}a and b). Both the uniaxial force and electric current were applied along this \textit{x}~axis. The angle between the magnetic field and \textit{x} axis is denoted as $\phi_{ab}$.

\begin{figure}
\begin{center}
\includegraphics[width=17.5cm]{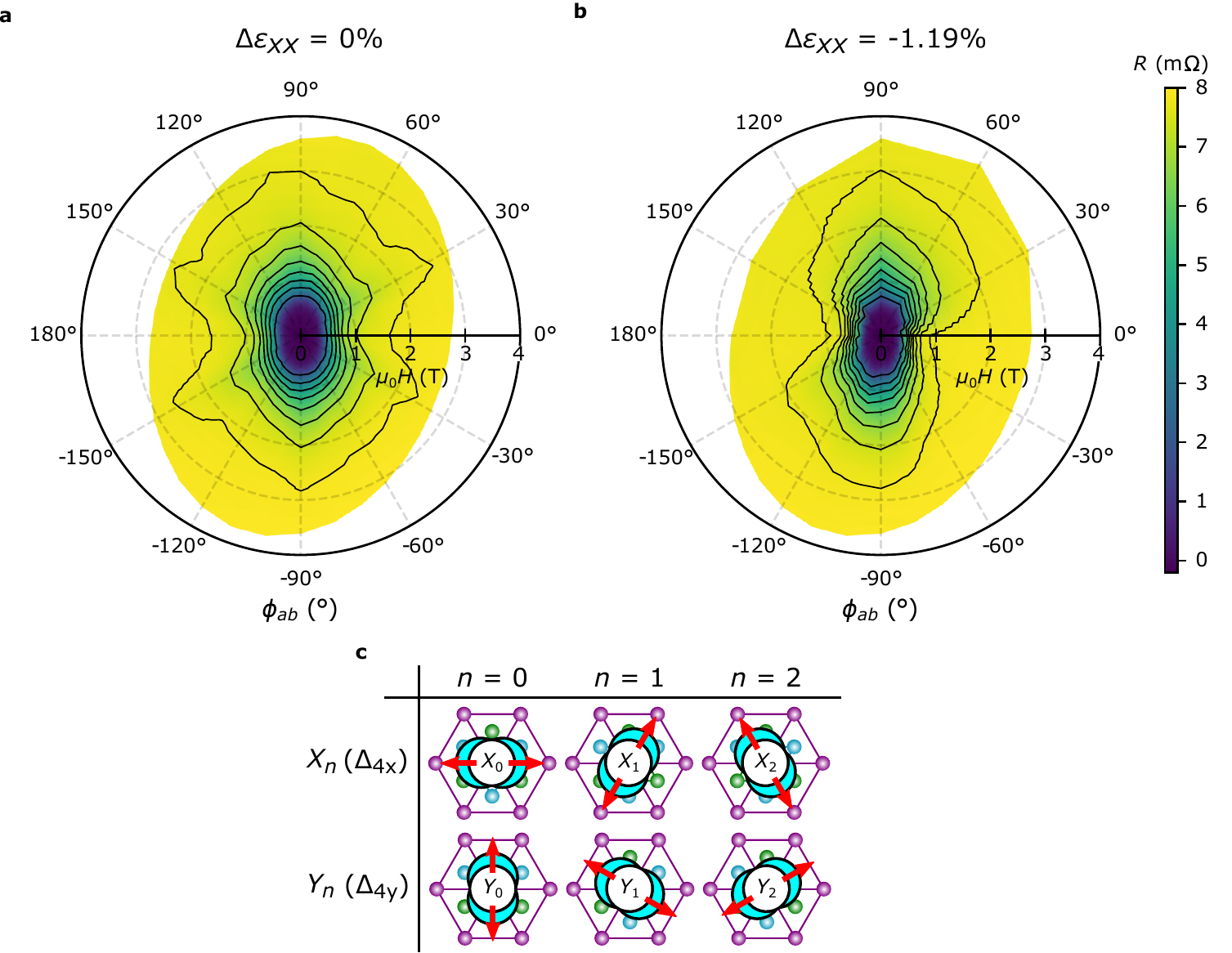}
\end{center}
\caption{
{\bf Disappearance of nematic superconducting domains in \sbs\ under compressive strain.}
{\bf a} and {\bf b}, Color polar plot of magnetoresistance for $H \parallel ab$ measured at the relative strains of $\Delta\varepsilon_{xx} = 0$ ({\bf a}) and $-1.19$\% ({\bf b}) with \SI{250}{\micro\ampere} applied current and 2.2~K.
The light-green regions extending along $\pm30$ and $\pm 150$\deg\ in {\bf a} indicate existence of nematic subdomains, which substantially disappears under applied strain ({\bf b}).
The contours are drawn from \SI{0.5}{m\ohm} to \SI{7.5}{m\ohm} in steps of \SI{1}{m\ohm}.
{\bf c}, Table of the 6 possible nematic superconducting states that can exist in the sample as domains. $X_{n}$ and $Y_{n}$ ($n=0,1,2$) domains exhibit $\Delta_{4x}$ and $\Delta_{4y}$ states with the large \Hcc\ along one of the \textit{a} axes ($\phi_{ab} = (60n)$\deg) and \textit{a}* axes ($\phi_{ab} = (90 + 60n)$\deg), respectively, as indicated with the red arrows. The crystal structure in the \textit{ab} plane of \bs~is shown with the schematic superconducting wave function in its center. The thickness of the blue crescent depicts the superconducting gap amplitude.
\label{fig2}
}
\end{figure}

%\textbf{[Raw data R(H) without strain]} 
In Figs.~\ref{fig1}c, \ref{fig2}a, and \ref{fig2}b, we present the magnetoresistance at 2.2~K for various \phiab. First, focus on the data with the relative strain \De\ of 0\%, i.e. zero applied voltage to the piezo stacks (black curves in Fig.~\ref{fig1}c), corresponding to the actual strain of around $+0.10\%$ (tensile) due to the thermal-contraction difference of the sample and strain cell (Methods). Clearly, superconductivity is more stable for $\phi_{ab} = -90$\deg~($H\parallel -y$) than $0$\deg~($H\parallel x$), resulting in a prominent two-fold upper critical field \Hcc, which is indicative of the nematic superconductivity~\cite{Pan2016.SciRep.6.28632, Du2017.SciChinaPhysMechAstron.60.037411}. This observed anisotropy $H\subm{c2} \parallel -y> H\subm{c2} \parallel x$ is consistent with the \Dy~state with the SC gap larger along \textit{y}~\cite{Tsutsumi2016.PhysRevB.94.224503}, which is schematically shown as the \Y{0}~state in Fig.~\ref{fig2}c. Interestingly, additional six-fold behavior emerges at the onset of the SC transition between 1 and 2~T, as clearly visible by the green region extending along $\phi_{ab} = -30$\deg\ or $+30$\deg\ in Fig.~\ref{fig2}a (See Supplementary Fig.~\ref{fig:Supp_RvsB_allAngles} for raw data). This six-fold component indicates that the sample contains minor parts exhibiting large \Hcc\ along $\phi_{ab}= \pm 30$\deg, namely the \Y{1}~and \Y{2}~domains (\Dy) in Fig.~\ref{fig2}c with their gap maxima along the $\pm30$\deg\ directions. 

%\textbf{[Raw data R(H) with strain, strain-control of nematic SC]} 
Next, let us focus on the data under applied strain of $\Delta\varepsilon_{xx}= -1.19$\% (green curves in Fig.~\ref{fig1}c) corresponding to the actual compressive strain of around $\varepsilon_{xx} \simeq -1.1$\%; the largest measured compressive strain in the elastic limit (see Supplementary Note~\ref{sec:Supp_RvsT_Irreversible}). Notably, the magnetoresistance at the SC transition is substantially altered, marking the first in-situ uniaxial-strain control of nematic superconductivity. More specifically, the SC transition becomes sharper with strain except near $\phi_{ab} = -90$\deg~($H\parallel y$). Moreover, comparing the color plots in Figs.~\ref{fig2}a and b, we can notice that the weak six-fold SC onset due to domains, seen in the $\Delta\varepsilon _{xx} = 0$ data, is substantially reduced by the applied strain. Thus the primary effect of the compressive uniaxial strain is to suppress the minor nematic domains.

\begin{figure}
\begin{center}
\includegraphics[width=12cm]{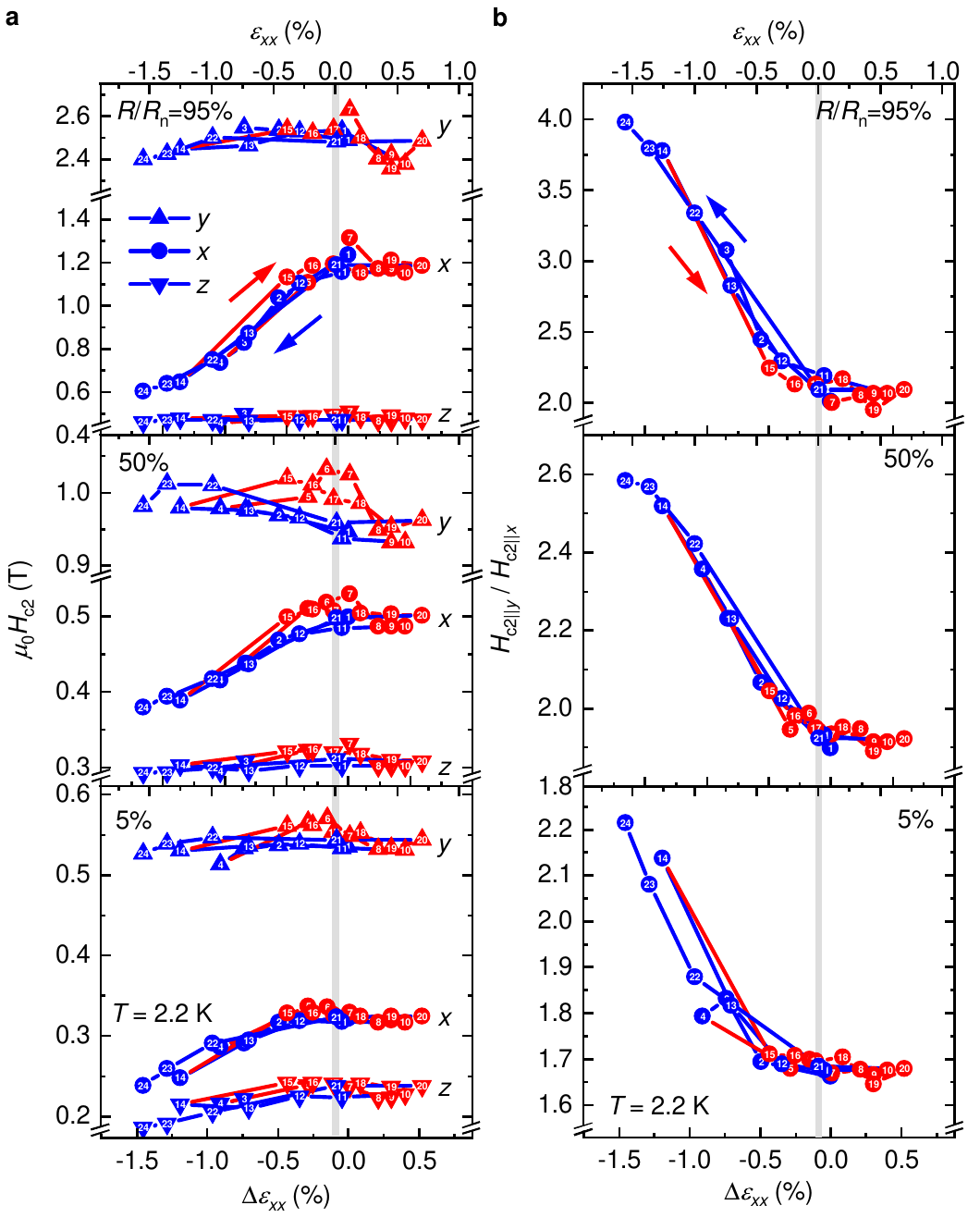}
\end{center}
\caption{
{\bf Reversible uniaxial-strain control of the nematic superconductivity.}
{\bf a}, Upper critical field (\Hcc) at 2.2~K along the \textit{a} axis ($x$; \phiab=0\deg; $\bullet$), \textit{a}* axis ($y$; \phiab=-90\deg; $\blacktriangle$), \textit{c} axis ($z$; $\blacktriangledown$), as a function of the relative strain $\Delta\varepsilon_{xx}$ induced by an applied voltage to the piezostacks. 
{\bf b}, In-plane \Hcc\ anisotropy $H\subm{c2\parallel \textit{y}} / H\subm{c2\parallel \textit{x}}$ as a function of strain.
At the top of each panel, the estimated actual strain $\varepsilon_{xx} \simeq \Delta\varepsilon_{xx} + 0.1\%$ is indicated (see Methods) and the gray region illustrates the possible range in the actual zero strain.
The numbers in the top corner of each sub-panel indicates the criteria used for determining \Hcc~(see Methods) %Supplementary Note~\ref{Supp_RvsH_criteria}).
The numbers in the data points indicate the order of the measurements.
The blue and red data points indicate the cases that the measurement was performed after a decrease and increase in strain, respectively.
Here, the \Hcc~anisotropy varies systematically with external strain.
\label{fig3}
}
\end{figure}

%\textbf{[\Hcc~vs. strain]} 
From the magnetoresistance data, we defined \Hcc\ as the field where the resistance $R(H)$ divided by the normal-state resistance $R\subm{n}$ reaches various criterion values (Methods; Supplementary Note~\ref{sec:Supp_RvsH_criteria}).
In the strain dependence of \Hcc\ (Fig.~\ref{fig3}), there is a high reproducibility among measurement cycles within the present strain range, manifesting that strain response is repeatable and thus our sample is in the elastic deformation regime. Reproducibility across samples has also been demonstrated (see Supplementary Note~\ref{sec:Supp_RvsH_RvsT_S2}).

Comparing data for various field directions, we can see that $H\subm{c2}\parallel x$ largely reduces under strain, attributable to the disappearance of minor nematic SC domains. In contrast, \Hcc\ along the \textit{y} and \textit{z} axes (Fig.~\ref{fig3}), as well as the zero-field \Tc~(Supplementary Fig.~\ref{fig:Supp_TcvsDexx}), is only weakly affected by strain, with small decreasing trend under compression. 

\begin{figure}
\begin{center}
\includegraphics[width=15.5cm]{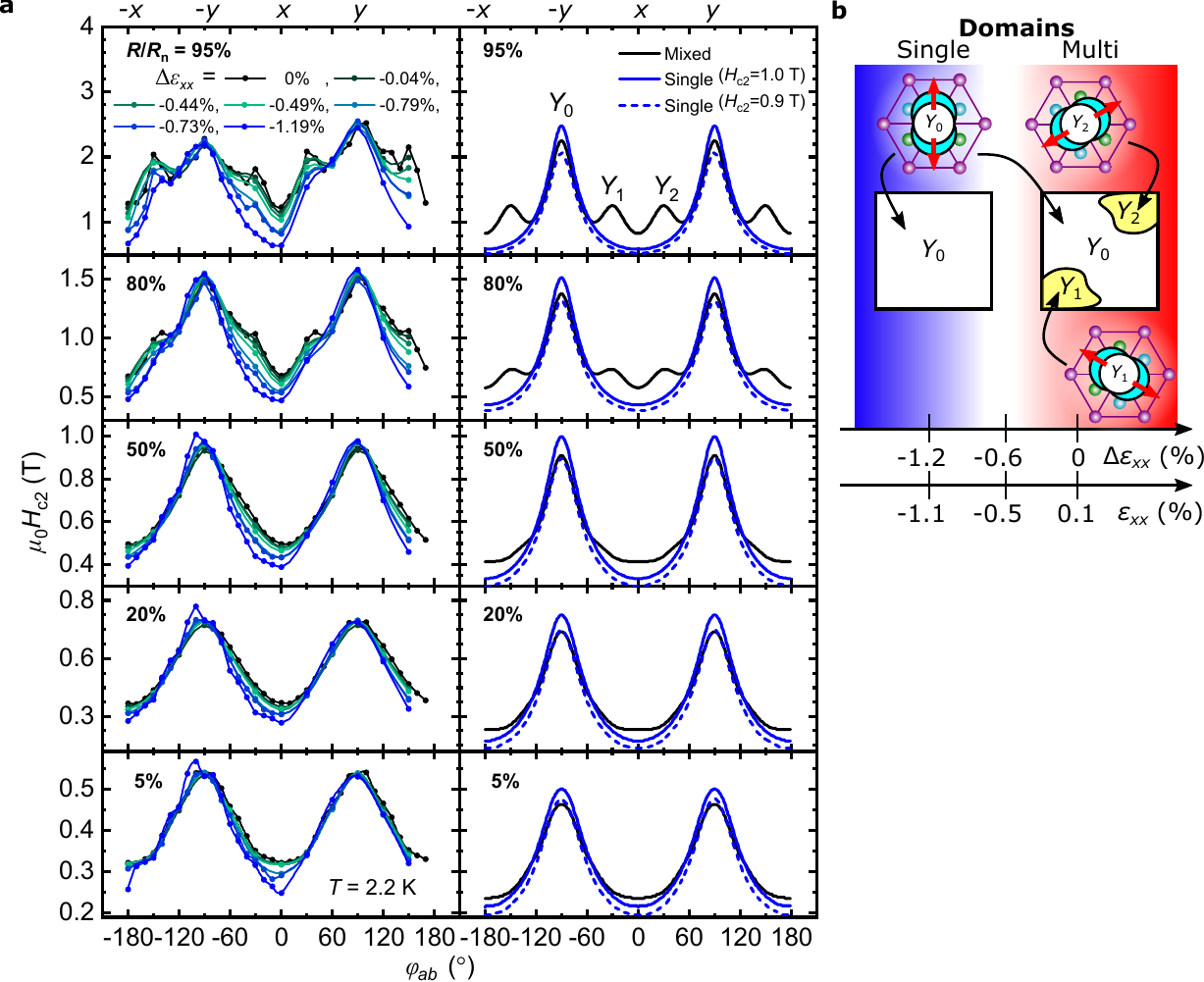}
\end{center}
\caption{
{\bf Evidence for the control of nematic superconducting domains with strain.}
{\bf a}, In-plane field angle \phiab\ dependence of the upper critical field \Hcc\ at 2.2~K for various criteria (see Methods), which are indicated with a number in the top-left corner of each sub-panel. The curves colored from black to blue are in the order of increasing compressive strain; the numbers in the legend indicate the value of $\Delta\varepsilon_{xx}$. The results of a simulation are shown in the second column (see Methods) capturing key features of the observation.
{\bf b}, Schematic showing spatial configurations of nematic superconducting domains controlled in-situ by the uniaxial strain in our experiment. The yellow regions are the minor domains (\Y{1} and \Y{2}), which are suppressed by the application of compressive strain, as evidenced by the changes in the $H\subm{c2}(\phi_{ab})$ curves.
\label{fig4}
}
\end{figure}

%\textbf{[\Hcc~vs. phi, simulation]} 
The strain control of the nematic subdomains is more evident in the \Hcc(\phiab) curves in Fig.~\ref{fig4}a. Notice that \Hcc\ defined with higher values of $R/R\subm{n}$ is more sensitive to existence of nematic sub domains. In addition to the prominent two-fold anisotropy with maxima at $\phi_{ab} = \pm 90$\deg\ (\Y{0}~domain) seen in all criteria, \Hcc\ with the 95\% or 80\% criteria exhibit additional 4 peaks located at $\phi_{ab} = \pm 30$\deg\ and $\pm 150$\deg\ for low \De, due to the existence of \Y{1}~and \Y{2}~domains. These peaks are suppressed with increasing \De, indicating the disappearance of the minor \Y{1}/\Y{2}~domains. 
In contrast, in \Hcc\ with lower criteria, the additional peaks are absent because the sample resistivity near the zero resistance state is mostly governed by the domain with the highest volume fraction.
Nevertheless, even for \Hcc\ with the lower criteria (e.g. $R/R_n = 5$\%), there is noticeable strain dependence near $\phi_{ab} = 0$\deg. 
This dependence is also attributable to the domain change by comparison with a model simulation explained next. 

In this simulation, we assume a network consisting of many \Y{0}~domains and one of each \Y{1}/\Y{2}~domain and calculate the net resistance under magnetic fields (Methods). 
Then \Hcc\ with various criteria is evaluated from the calculated resistivity curves. Under strain, the minor domains are assumed to change into \Y{0}~domains. As shown in Fig.~\ref{fig4}, the simulation reproduces all the observed features described above, even without any change of \Hcc~and in-plane \Hcc\ anisotropy in each domain. This leads one to infer that the observed behavior is almost solely explained by the change of the nematic SC subdomains. We find that setting slightly smaller \Hcc~(ca. decrease of 10\%; the broken curves in Fig.~\ref{fig4}a) of the \Y{0} domain under strain gives a better match with the experimental data. 

%\textbf{[Summary of experimental findings]} 
Summarizing our findings, we succeeded in repeatable in-situ uniaxial-strain control of nematic SC domains, covering pre-existing tensile regime to the compressive regime. The primary effect of the increasing compressive strain is the suppression of minor \Y{1}/\Y{2}~domains, well reproduced by a simple model simulation. 
Other properties are rather insensitive to the strain, but there are decreasing trends in \Tc\ as well as \Hcc\ of the main domain under compression.

%\section{[Discussion]} %Comment out

%\textbf{[Comparison with GL theory]} 
The coupling between the nematic superconductivity and uniaxial strain has been proposed using the Ginzburg-Landau (GL) formalism~\cite{Venderbos2016.PhysRevB.94.094522, How2019}. The strain couples to the nematic superconductivity through the free energy $F_{\varepsilon} = g \left[ (\varepsilon_{xx}-\varepsilon_{yy})(|\eta_x|^2 - |\eta_y|^2) + 2\varepsilon_{xy} (\eta_x\eta_y^\ast + \eta_y\eta_x^\ast) 
\right]$, where $g$ is the coupling constant, $\eta_x$ and $\eta_y$ are the amplitudes of the $\Delta_{4x}$ and $\Delta_{4y}$ components. This relation indicates that uniaxial $\varepsilon_{xx}$ strain prefers one of the \Dx\ or \Dy\ states, depending on the sign of $g$. 
If a pre-existing symmetry breaking field exists, the nematic SC order parameter is initially fixed to the pre-existing field direction but eventually the state most favored by the external strain direction will be chosen with increasing strain. 
This is true even when the pre-existing field and external strain have a finite angle, as in the case for the \Y{1}\ or \Y{2}\ domains: the nematicity gradually rotates toward the external strain (see Supplementary Fig.~\ref{fig:Supp_NematicSwitching}).
However, in these phenomenological theories, the sign of $g$ remains arbitrary and should be determined based on experiments.  Our result, a multi-domain sample driven to a mono-domain \Dy\ state  by $\varepsilon_{xx}<0$ as shown in Fig.~\ref{fig3}c, indicates that $g$ is negative, an important step toward modeling of the nematic SC phenomenon. Moreover, this negative $g$ provides a crucial constraint to realistic microscopic theories on the pairing mechanism. Such model should explain the observed weak sensitivity of \Tc\ on $\varepsilon_{xx}$. For example, a proposed odd-parity fluctuation model making use of phonons dispersing along the $k_{z}$~direction\cite{wan2014turning,PhysRevB.90.184512,PhysRevB.96.144504,Wang2019} can be compatible with our observation, since such $k_{z}$~phonons should be less sensitive to the in-plane distortions. 

%\textbf{[Comment on some "inconsistency" with GL theory]} 
Coming back to the GL theories, they predict that \Tc\ linearly increases with increasing strain in either tensile or compressive directions, accompanied by a kink in $T\subm{c}(\varepsilon_{xx})$ at the strain where the nematic state changes between \Dx\ and \Dy\cite{Venderbos2016.PhysRevB.94.094522, How2019}. This prediction, at first glance, seems to be inconsistent with our decreasing trend of \Tc\ with increasing $|\varepsilon_{xx}|$. However, we should note that \Tc\ of doped \BS\ decreases under hydrostatic pressure, i.e. under isotropic strain~\cite{Nikitin2016.PhysRevB.94.144516}. This effect is not taken into account in the above mentioned GL free energy, which couples only to the anisotropic strains. In the actual experiments, a combination of the increasing and decreasing trends in \Tc\ due to anisotropic and isotropic strains, respectively, is observed. If the latter is relatively stronger, the observed small decrease of \Tc\ by compressive strain is explained. 
%In addition, if the sample is already in the tensile regime then $T\subm{c}(\varepsilon_{xx})$ should decrease with compressive strain. The calculation of thermally induced strain places our sample in the tensile regime, neglecting any offset due a pre-exiting symmetry breaking field, thus supporting this possibility as well for the decreasing trend in \Tc.
Moreover, the existence of multiple domains weakens the predicted kink in $T\subm{c}(\varepsilon_{xx})$, because each domain's $T\subm{c}(\varepsilon_{xx})$ curve convolves. This will result in a rounded kink, further obfuscating the linear behavior predicted from a mono-domain model.

To conclude, we provide the first experimental demonstration of uniaxial-strain control of nematic superconductivity in doped \BS. Firstly, suppressing minor domains while stabilizing the \Dy~state. Secondly, we determined the sign of the nematic coupling constant. These findings should provide bases toward resolving the open issues of this highly attractive superconductor. Additionally, this work points to possible engineering of topological nematic superconductivity by uniaxial strain.

\begin{methods}

% \section*{Sample preparation and characterization.}
\textbf{Sample preparation and characterization.} Single crystals of \sbs~(nominal $x = 0.06$) were grown from high-purity elemental of Sr chunk (99.99\%), Bi shot (99.9999\%), and Se shots (99.9999\%) by a conventional melt-growth method. The raw materials were mixed with a total weight of 5.0 g and sealed in an evacuated quartz tube. The tube was heated to 1223 K and kept for 48 hours with intermittent shaking to ensure the homogeneity of the melt. Then it was cooled slowly to 873 K at a rate of 4 K/h and finally quenched into ice water. It is worth pointing out that quench is essential for obtaining superconducting samples with high shielding fraction. The sample used here was cut from a large shiny crystal by wire saw, and the size is 4~mm (length) $\times$ 0.53~mm (width) $\times$ 0.5~mm (thickness; along the \textit{c} axis) with the longest dimension along the $a$ axis. 

%\section*{Strain cell and sample mounting.}
\noindent\textbf{Strain cell and sample mounting.} We constructed a custom-made piezoelectric-based uniaxial strain cell (Ref. \citenum{JAPKostylev2019}), based on the design of Ref. \citenum{Hicks2014RSI}. The bar-shaped sample was mounted between two anvils by a strong epoxy (Stycast 2850FTJ, Henkel Ablestik Japan Ltd.). The anvils can apply compressive or tensile strain on the sample by applying a positive voltage on the inner or outer piezo stacks, respectively. Thus the strain was applied parallel to the $a$ axis, as shown in Fig.~1. The maximum applied voltage range for each piezo stack was $-400$~V to 600~V corresponding roughly to \SI{-13}{\micro\meter} to \SI{20}{\micro\meter} length changes of the piezo stacks used (P-885.11, PI) at cryogenic temperatures. A parallel-plate capacitor was mounted on the anvils to track the distance between the two plates by measuring the capacitance using a capacitance bridge (2500A, Andeen-Hagerling). The strain was then determined by the displacement divided by the exposed sample length, which was $1.14\pm 0.05$~mm in this study.

\noindent\textbf{Estimation of the thermally-induced strain.} The effect of thermal contraction of the sample and the strain cell should be taken into consideration. Because the materials used in the strain cell are placed symmetrically between the compressive and tensile arms, the thermal strain on the sample originates only from the asymmetric part\cite{JAPKostylev2019}; on the compressive arm, the sample with the length $L\subm{sample}$ of 1.14~mm is placed, but on the tensile arms there are Ti blocks. This 1.14-mm length Ti part shrinks less than the sample, resulting in a tensile strain to the sample after cooling down from the epoxy curing temperature (around 350~K). 
The shrinkage of the sample $\Delta L\subm{sample}/L\subm{sample}$ is evaluated as $[a(4\text{ K}) - a(350\text{ K})]/ a(350\text{ K}) = -0.36$\%.
Here, we used the lattice constants of \BS\ reported in Ref.~\citenum{Chen2011}.
We note that $a(4\text{ K}$) and $a(350\text{ K}$) is estimated by a linear extrapolation because Ref.~\citenum{Chen2011} reports $a$ values only between 10~K and 270~K.
For Ti, the shrinkage $\Delta L\subm{Ti}/L\subm{Ti}$ is evaluated to be $-0.23$\% by integrating the linear thermal expansion coefficient between 4~K and 350~K reported in Ref.~\citenum{COWAN1968155}. The thermal expansion coefficient at 4~K and 350~K were obtained after linearly extrapolated.
Thus, the thermally-induced strain to the sample is tensile and $(\Delta L\subm{Ti} - \Delta L\subm{sample})/L\subm{sample} = +0.13$\% considering $L\subm{Ti} = L\subm{Sample}$.
In addition, because of the stiffness of the component materials, in particular the epoxy, the actual strain transmitted to the sample may be reduced by roughly 56\%~\cite{JAPKostylev2019}. Thus, the value +0.13\% should be considered as the upper bound, and the lower bound should be $0.13\%\times 0.56 = 0.07$\%.
To conclude, by taking the average, the thermally-induced strain is evaluated to be $0.10 \pm 0.03$\%: the actual strain $\varepsilon_{xx}$ is given as $\varepsilon_{xx} \simeq \Delta\varepsilon_{xx} + 0.1\%$, where $\Delta\varepsilon_{xx}$ is the strain applied relative to the situation of zero applied voltage to the piezo stacks.
\noindent\textbf{Resistivity measurement.} Sample resistivity was measured by four-terminal sensing: we applied a DC current using a current source (6221, Keithley Instruments) to the two outer wires and measure the resultant voltage by a nanovoltmeter (2182A, Keithley Instruments) on the inner two wires. To subtract the voltage offset, we use the ``Delta Mode'' of the combined operation of these instruments: the polarity of the current was periodically changed to acquire only the voltage component that is dependent on the current. Au wire (\SI{20}{\micro\meter} diameter) were directly connected to the \textit{ac} surface of the sample by Ag paint (4929N, Du Pont). To improve the mechanical stability of the wires, the Au wires were anchored onto the $ab$ surface by Ag epoxy (H20E, EPOTEK), which has been confirmed to be electrically insulating to the sample. The four contacts were equispaced by about 0.2~mm. The contact resistance was on the order of 100~$\Omega$ at room temperature.

%\section*{Temperature and magnetic-field control.}
\noindent\textbf{Temperature and magnetic-field control.} We used a $^3$He/$^4$He dilution refrigerator (Kelvinox 25, Oxford Instruments) to cool down the sample. It was inserted into the vector magnet described below. The lowest temperature achievable is roughly 80~mK, well below the superconducting transition temperature of $\sim2.8$~K. The temperature was measured using a resistive thermometer (Cernox, Lakeshore) and a resistance bridge (AVS-47, Picowatt). A 350-$\Omega$ strain gauge (KFG-1-350-C1-16, KYOWA), that was used as a heater for temperature control, was mounted close to the strain cell.

We applied the magnetic field using a vector-magnet system\cite{Deguchi2004}, which consists of two orthogonal superconducting magnets (pointing in the vertical and horizontal directions in the laboratory frame) inside a dewar that rests on a horizontal rotation stage.  This system allows us to direct the magnetic field accurately in any direction in space while the refrigerator, as well as the sample, is fixed. The superconducting magnets can apply fields up to 3 T (vertical) and 5 T (horizontal). The magnetic field can be controlled with a resolution of 0.1 mT. 
%Thus an angular resolution for the polar angle of the field depends on the magnitude of the applied fields but typically around xxx degrees. 
The precision of the horizontal rotation of the helium dewar is 0.001\deg, with negligible backlash. The strain cell was fixed with a sample mounted, so that the \textit{a} axis is roughly along the vertical direction in the laboratory frame. The precise directions of the crystalline axes with respect to the laboratory frame are determined by making use of the anisotropy in \Hcc. Once the directions of the crystalline axes are determined, we can rotate the magnetic field within the sample frame. All field angle values presented in this Letter are defined in the sample frame.
Refer to Supplementary Note~\ref{sec:Supp_EulerTransformExample} and Fig.~\ref{fig:Supp_EulerTransformExample} for the detailed mathematical explanation for the vector transformations and a demonstration of the field alignment.
In addition, see Supplementary Note~\ref{sec:Supp_R_Phi_Theta} for the rationale behind the choice of temperature and magnetic field value for the alignment.

%\section*{Data analysis.}
\noindent\textbf{Evaluation of \Hcc.} The upper critical field \Hcc~was evaluated by the value of the magnetic field at which the sample resistivity reaches a certain percentage of the normal-state resistivity. If the resistivity value falls in between two data points, then \Hcc~is determined by using linear interpolation. For a more detailed methodology see Supplementary Note~\ref{sec:Supp_RvsH_criteria}.
Temperature dependence of \Hcc\ and its anisotropy is given in Supplementary Note~\ref{sec:Supp_Hc2vsT_AnivsT}.

%\section*{Model simulation.}
\noindent\textbf{Model simulation.} The experimental \Hcc(\phiab) curve shows three peaks indicating three nematic domains. Thus, we simulated the \Hcc\ behavior of multi and single domain samples by considering an electrical circuit consisting of a network of resistive elements representing the three possible nematic SC domains. For the simulation shown in the main text, the circuit is assumed to be a 3D network (see Supplementary Fig.~\ref{fig:Supp_3DCircuit}) of 12 elements to model the situation that current passes from end to end through a 3D distribution of domains. For the multi-domain simulation corresponding to the non-strained sample, the 12 elements are divided into 10 \Y{0} nematic domains, and one of each \Y{1} and \Y{2} domains. The exact positions are described in Supplementary Note~\ref{sec:Supp_3DCircuit}. For the single-domain simulation corresponding to the highly compressed sample, all the 12 elements are assumed to be \Y{0}\ domains.
The calculation of \Hcc~is done as follows: firstly, for a fixed $H$ and \phiab, resistivity of each circuit element is calculated from an empirical relationship among resistance, applied magnetic field $H$ and field direction \phiab, by taking into account \Hcc\ anisotropy of each domain (see Supplementary Note~\ref{sec:Supp_3DCircuit} for details). Secondly, the total circuit resistance of the network $R\subm{total}$ is calculated. The first and second step is iterated while varying $H$ and \phiab, to obtain the $H$ dependence of $R\subm{total}$ for each $\phi_{ab}$. Lastly, \Hcc\ at \phiab\ is determined from the $R\subm{total}(H)$ curve at \phiab\ using the same method as that used for the experimental data analysis (see Supplementary Note~\ref{sec:Supp_RvsH_criteria}).

\end{methods}

%\stopcoloring %Comment out

\bibliography{References}

\section*{Acknowledgments}
The authors acknowledge H.-H.~Wen, J.~Schmalian, P.~T.~How, S.-K.~Yip, V.~Kozii, and J. W. F.~Venderbos, H.-S.~Xu, and G.~Mattoni for valuable discussions.
We also acknowledge C.~W.~Hicks, M.~E.~Barber, A.~Steppke, F.~Jerzembeck, and A.~P.~Mackenzie for sharing their knowledge into the construction of a strain cell.
This work was supported by JSPS Grant-in-Aids for Scientific Research on Innovative Areas on ``Topological Materials Science''  (KAKENHI JP15H05851, JP15H05852, JP15H05853, JP15K21717), by JSPS Grant-in-Aid KAKENHI 17H04848, and by the JSPS Core-to-Core program.
The work at Cologne was funded by the Deutsche Forschungsgemeinschaft (DFG, German Research Foundation) - Project number 277146847 - CRC 1238 (Subproject A04).

\section*{Author contributions}

This study was designed by I.K., S.Y., and Y.M.;
I.K. performed resistivity measurements and analyses, with the assistance of S.Y. and the guidance of Y.M.;
Z.W. and Y.A. grew single crystalline samples and characterized them.
The uniaxial strain cell was designed and constructed by I.K.;
The manuscript was prepared mainly by I.K. and S.Y., based on discussion among all authors.

\section*{Competing financial interests}

All authors declare there is no competing interests regarding this work.

% \section*{Figures}
% \input{Figures.tex}

\clearpage
\appendix
%UN-COMMENT NEXT 3 LINES IF USING EXTERNAL DOCUMENT
%\documentclass{nature_fig}
%\usepackage{Preamble}
%\myexternaldocument{main}

%Supplementary Specific
\makeatletter
\renewcommand{\section}{\@startsection {section}{1}{0pt}{0pt}{1pt}{\reset@font\Large\bfseries}}
\renewcommand{\subsection}{\@startsection {subsection}{1}{0pt}{0pt}{1pt}{\reset@font\large\bfseries}}
\renewcommand{\theequation}{S\arabic{equation}}
\setcounter{equation}{0}
\renewcommand{\thefigure}{S\arabic{figure}}
\setcounter{figure}{0}
\renewcommand{\thetable}{S\arabic{table}}
\setcounter{table}{0}
\renewcommand{\thepage}{S\arabic{page}}
\setcounter{page}{1}
\renewcommand{\thesubsection}{S\arabic{subsection}}
\setcounter{equation}{0}
\renewcommand{\figurename}{Supplementary Figure}
\renewcommand{\tablename}{Supplementary Table}
\spacing{2}
\setlength{\parskip}{10pt}
\renewcommand{\refname}{\setlength{\parskip}{12pt}Supplementary Reference}
\renewcommand{\bibnumfmt}[1]{[S{#1}]}
%\bibpunct{[S}{]}{,~S}{n}{}{,}
%\renewcommand{\citenumfont}[1]{S{#1}}
\makeatother

\setcounter{topnumber}{100}
\setcounter{bottomnumber}{100}
\setcounter{totalnumber}{100}
\renewcommand{\topfraction}{1.0}
\renewcommand{\bottomfraction}{1.0}
\renewcommand{\textfraction}{0.0}
\renewcommand{\floatpagefraction}{0.0}

% \sloppy

% \bibliographystyle{naturemag_hold-case_noURL}

% \title{Supplementary Information for\\[0.2em]
% Uniaxial-strain Control of Nematic Superconductivity in \sbs}

% %% Notice placement of commas and superscripts and use of &
% %% in the author list

% \author{Ivan~Kostylev$^{1\ast}$, Shingo~Yonezawa$^{1\ast}$, Zhiwei~Wang$^{2,3}$,
% Yoichi~Ando$^{2}$, Yoshiteru~Maeno$^1$}

% \begin{document}

% \maketitle

% \vspace{-0.5cm} 

% \begin{affiliations}
% {\small
%  \item Department of Physics, Graduate School of Science, Kyoto University, Kyoto 606-8502, Japan
%  \item Institute of Physics II, University of Cologne, K\"{o}ln 50937, Germany
%  \item Key Laboratory of Advanced Optoelectronic Quantum Architecture and
% Measurement, Ministry of Education (MOE), School of Physics, Beijing
% Institute of Technology, Beijing 100081, P. R. China
% }
% \end{affiliations}

% \vspace{-0.8em}

% \noindent
% $^{\ast}$e-mail: kostylev@scphys.kyoto-u.ac.jp, yonezawa@scphys.kyoto-u.ac.jp

% \vspace{-1.5em}

% \noindent \textit{Dated: August 1, 2019}

% \vspace{1em}

% \clearpage

\spacing{1.2}
%%%%%%%%%%%%%%%%%%%%%%%%%%%%%%%%%%%%%%%%%%%%%%%%%%%%%%%%

\section*{Supplementary Note}

%%%%%%%%%%%%%%%%%%%%%%%%%%%%%%%%%%%%%%%%%%%%%%%%%%%%%%%%
\subsection{Temperature dependence of resistivity at zero field}
\label{sec:Supp_RvsT_dExx}

\begin{figure}[htb]
\begin{center}
\includegraphics[width=11cm]{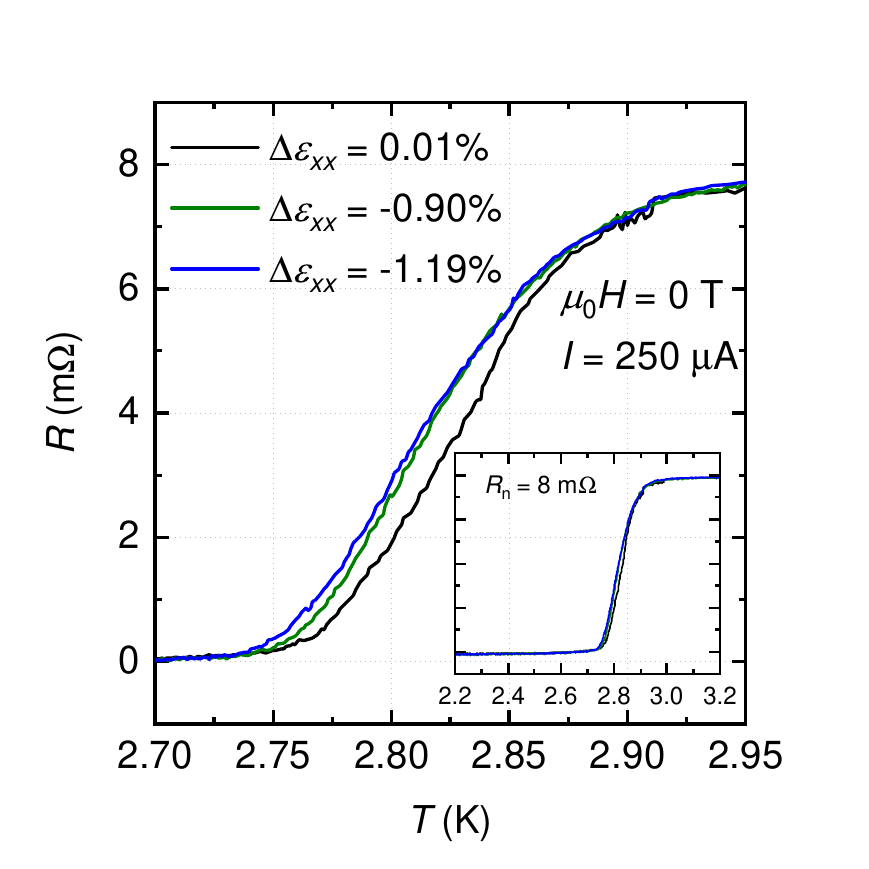}
\end{center}
\caption{
{\bf Temperature dependence of resistance at zero field and at various applied strain for \sbsn.}
These data were taken with the applied current \textit{I} of \SI{250}{\micro\ampere}. In the inset, larger temperature range is shown to demonstrate the nearly $T$-independent resistance in the superconducting ($T<2.7$~K) and normal state ($T>2.95$ K) regions. $R\subm{n} = 8$~m$\Omega$ is the resistance in the normal-state. The main figure and its inset share the same vertical axis scale.
\label{fig:Supp_RvsT_dExx}
}
\end{figure}

In Fig.~\ref{fig:Supp_RvsT_dExx}, we show the temperature dependence of the zero-field resistance of \sbsn\ at three different strain values. For these measurements, we used the applied current of \SI{250}{\micro\ampere}. The superconducting critical temperature (\Tc) defined as the mid-point of the transition is 2.83~K for zero relative strain \De, i.e. zero applied voltage to the piezo stacks. With increasing $|\Delta\varepsilon_{xx}|$, \Tc\ tends to decrease (See Supplementary Fig.~\ref{fig:Supp_TcvsDexx}). With compressive strain of \De$ = -1.19$\%, \Tc\ decreases weakly by about 12~mK.

\clearpage

%%%%%%%%%%%%%%%%%%%%%%%%%%%%%%%%%%%%%%%%%%%%%%%%%%%%%%%%
\subsection{Raw \textit{R} vs \textit{B} data for all in-plane field angles}
\label{sec:Supp_RvsB_allAngles}

\begin{figure}[htb]
\begin{center}
\includegraphics[width=12cm]{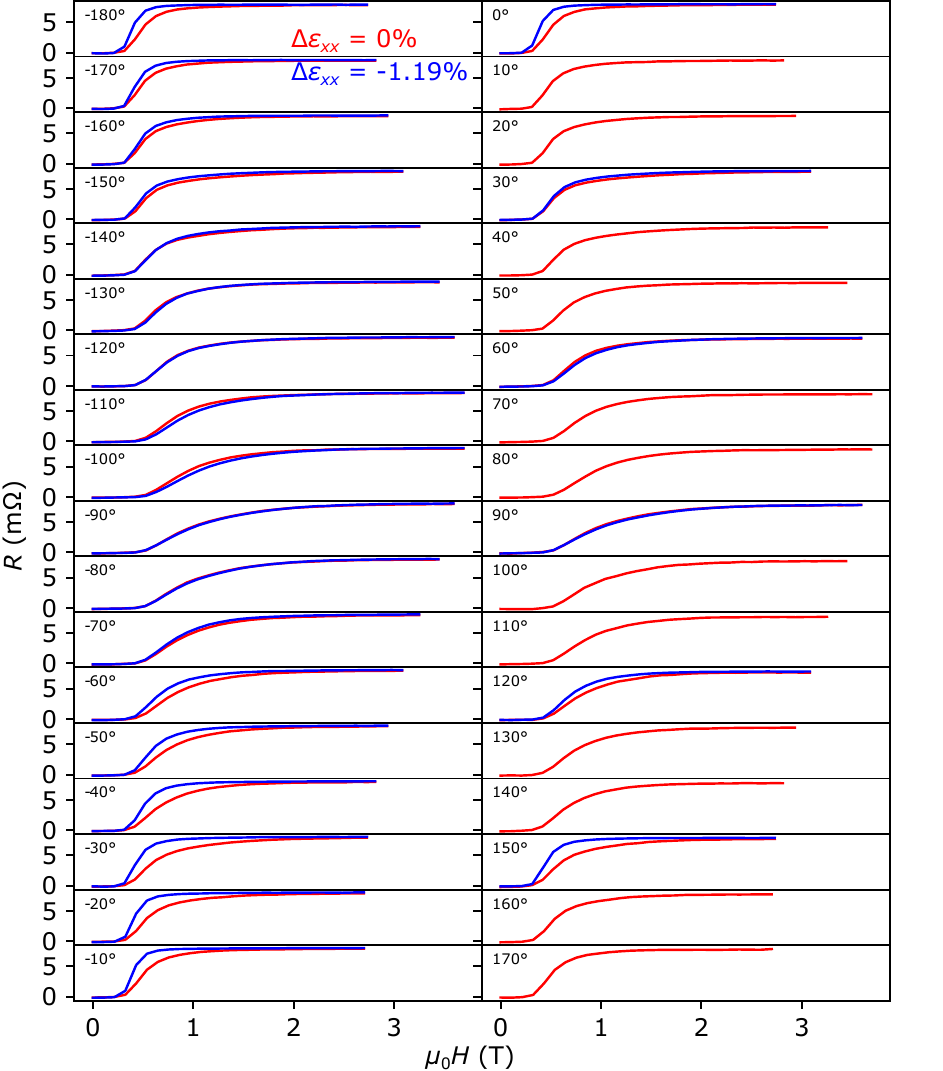}
\end{center}
\caption{
{\bf Magnetoresistance for various in-plane field angles.}
The red and blue curves correspond to zero applied strain ($\Delta\varepsilon_{xx} = 0$) and compressive strain ($\Delta\varepsilon_{xx} = -1.19$\%), respectively.
\label{fig:Supp_RvsB_allAngles}
}
\end{figure}

Here, we present a part of the raw magnetoresistance data, which are used to construct the color polar plot (Figs. 2a and b) and to evaluate \Hcc. In Fig.~\ref{fig:Supp_RvsB_allAngles}, we show the in-plane magnetoresistance of \sbsn\ for field angles in the range of $-180$\deg\ to 170\deg\ in steps of 10\deg. It is clear that with compressive strain of $\Delta\varepsilon_{xx} = -1.19$\%, \Hcc\ as well as the transition width decreases for $H \parallel x$ ($\phi_{ab}$ = 0\deg\ and $-180$\deg). Similar strain effect is also observed for the angles corresponding to the large \Hcc\ direction of the minor domains (i.e. $\phi_{ab} = \pm$30 and $\pm$150\deg).

\clearpage

%%%%%%%%%%%%%%%%%%%%%%%%%%%%%%%%%%%%%%%%%%%%%%%%%%%%%%%%
\subsection{Irreversible limit of the deformation}
\label{sec:Supp_RvsT_Irreversible}

\begin{figure}[htb]
\begin{center}
\includegraphics[width=16cm]{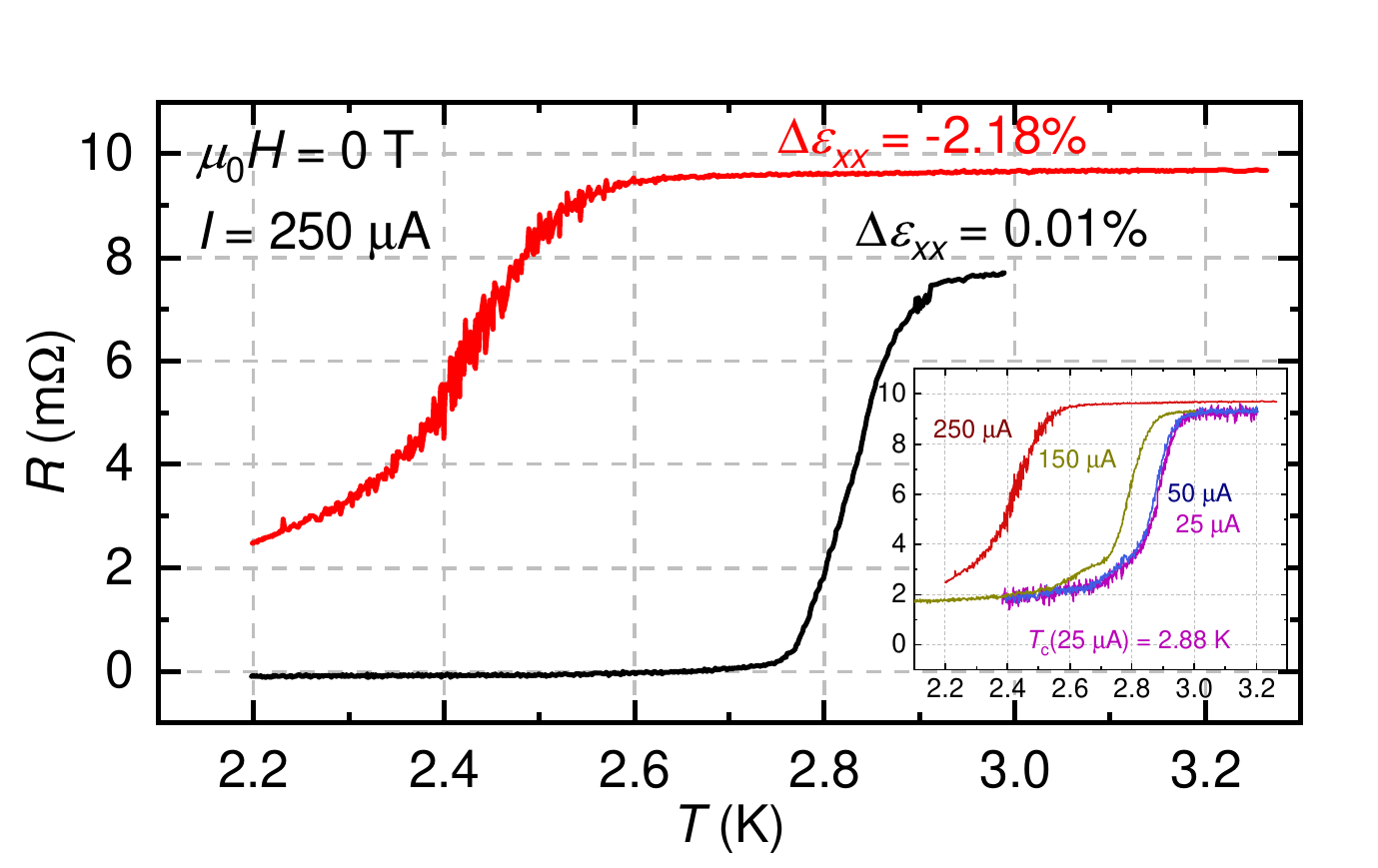}
\end{center}
\caption{
{\bf Zero-field resistance versus temperature before and after irreversible change.}
The black and red curves correspond to zero applied strain and large compressive strain, respectively. (inset) Resistivity curves measured with various currents after the irreversible change. The critical temperature returns close to the original value ($T\subm{c} = 2.88$~K) by decreasing applied current down to \SI{50}{\micro\ampere}.
\label{fig:Supp_RvsT_Irreversible}
}
\end{figure}

The strain range discussed in the Main Paper (i.e. $-1.5\% < \Delta\varepsilon_{xx} < +0.6\%$) is in the elastic deformation regime as evidenced by the reversible change of \Hcc\ (Fig.~3).
To support this claim, we sought for a border between the elastic and plastic regimes by applying stronger strains.
Indeed, as shown in Fig.~\ref{fig:Supp_RvsT_Irreversible}, we found that, after applying a sufficiently large compressive strain, the sample's electrical properties change irreversibly. After applying a large compressive strain of $\Delta\varepsilon_{xx} = -2.18$\%, the normal-state resistance increased by about \SI{2}{m\ohm} and \Tc\ shifts down to about 2.4~K ($\Delta T\subm{c} \sim -0.4$~K), although \Tc\ can be increased again by reducing the applied current from 250~$\mu$A down to 50~$\mu$A.
The shift of the normal state resistance persisted after releasing the strain.
The likely explanation is that formation of microcracks in the sample increases overall resistance and results in a Josephson-junction-like structure, which has much smaller critical current than bulk.

\clearpage

%%%%%%%%%%%%%%%%%%%%%%%%%%%%%%%%%%%%%%%%%%%%%%%%%%%%%%%%
\subsection{\Hcc\ criteria}
\label{sec:Supp_RvsH_criteria}

\begin{figure}[htb]
\begin{center}
\includegraphics[width=16cm]{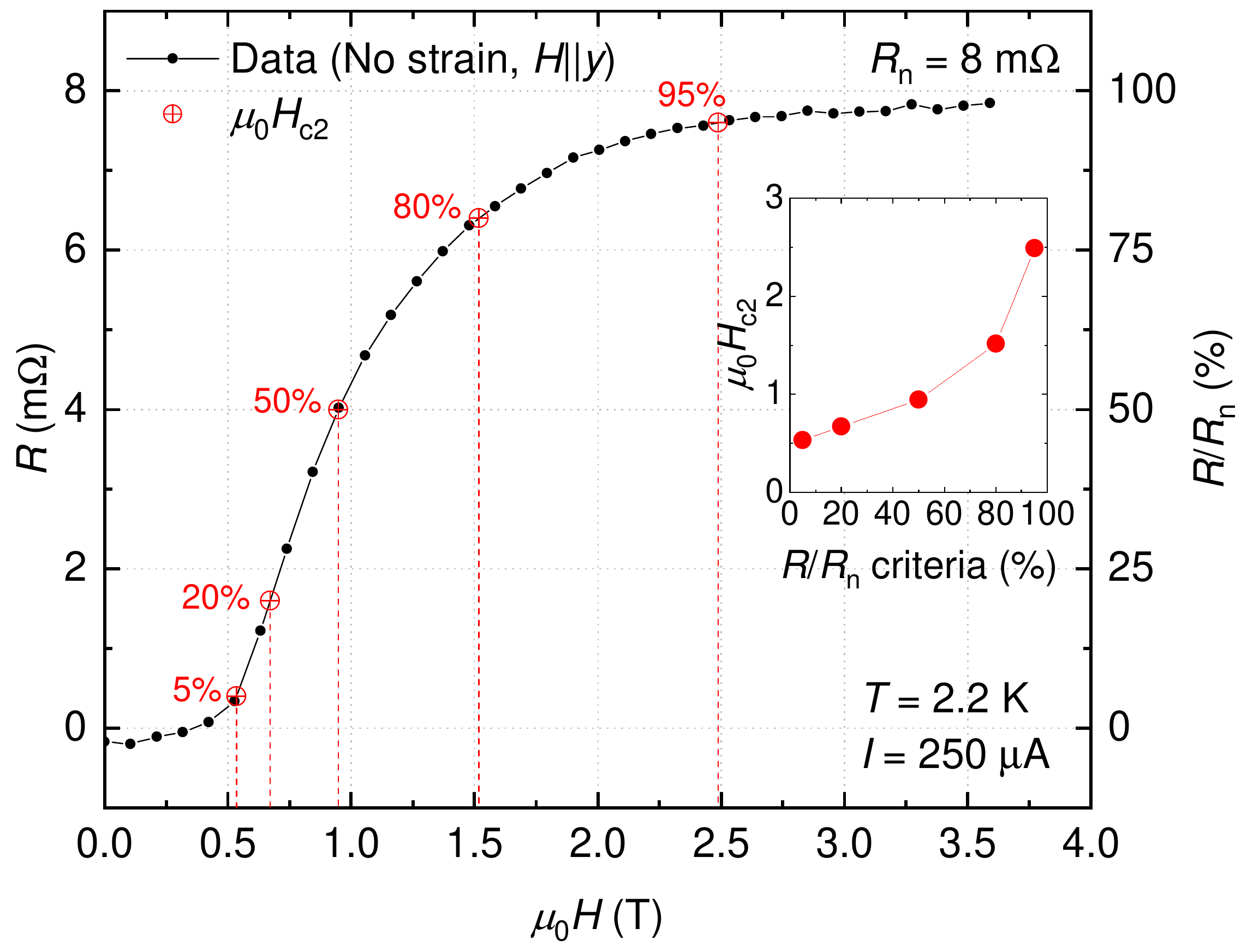}
\end{center}
\caption{
{\bf Methodology for determining the upper critical field \Hcc\ from magnetoresistance data.}
The data were taken under $\Delta\varepsilon_{xx} = 0$\% and for $H\parallel y$ ($\phi_{ab} = -90$\deg).
The red percentage labels indicate the $R/R\subm{n}$ criteria used, where $R\subm{n}$ is the normal-state resistance. The red circle and the corresponding vertical dotted line indicate the resistance value at the criteria and the corresponding field value (to be used as \Hcc), respectively. (Inset) \Hcc\ as a function of the $R/R\subm{n}$ criteria.
\label{fig:Supp_RvsH_criteria}
}
\end{figure}

In order to explain \Hcc\ evaluated in this work, we show in Fig.~\ref{fig:Supp_RvsH_criteria} an example how \Hcc\ is determined from magnetoresistance curves using the criteria in the ratio between the resistance $R$ and its normal-state value $R\subm{n}$. The process is to first decide on a criterion value of $R/R\subm{n}$, e.g. 50\%. Then the $R(H)$ curve is linearly interpolated in-between points to determine the precise value of the magnetic field at which $R/R\subm{n}$ reaches the criterion value: this value is taken to be \Hcc\ of that criterion. In this work, we employ criteria $R/R\subm{n}$ ranging from 5\% to 95\% to carefully examine the strain effect on nematic superconductivity.

\clearpage

%%%%%%%%%%%%%%%%%%%%%%%%%%%%%%%%%%%%%%%%%%%%%%%%%%%%%%%%
\subsection{Reproducibility of the strain control of nematic superconductivity}
\label{sec:Supp_RvsH_RvsT_S2}

\begin{figure}[htb]
\begin{center}
\includegraphics[width=14cm]{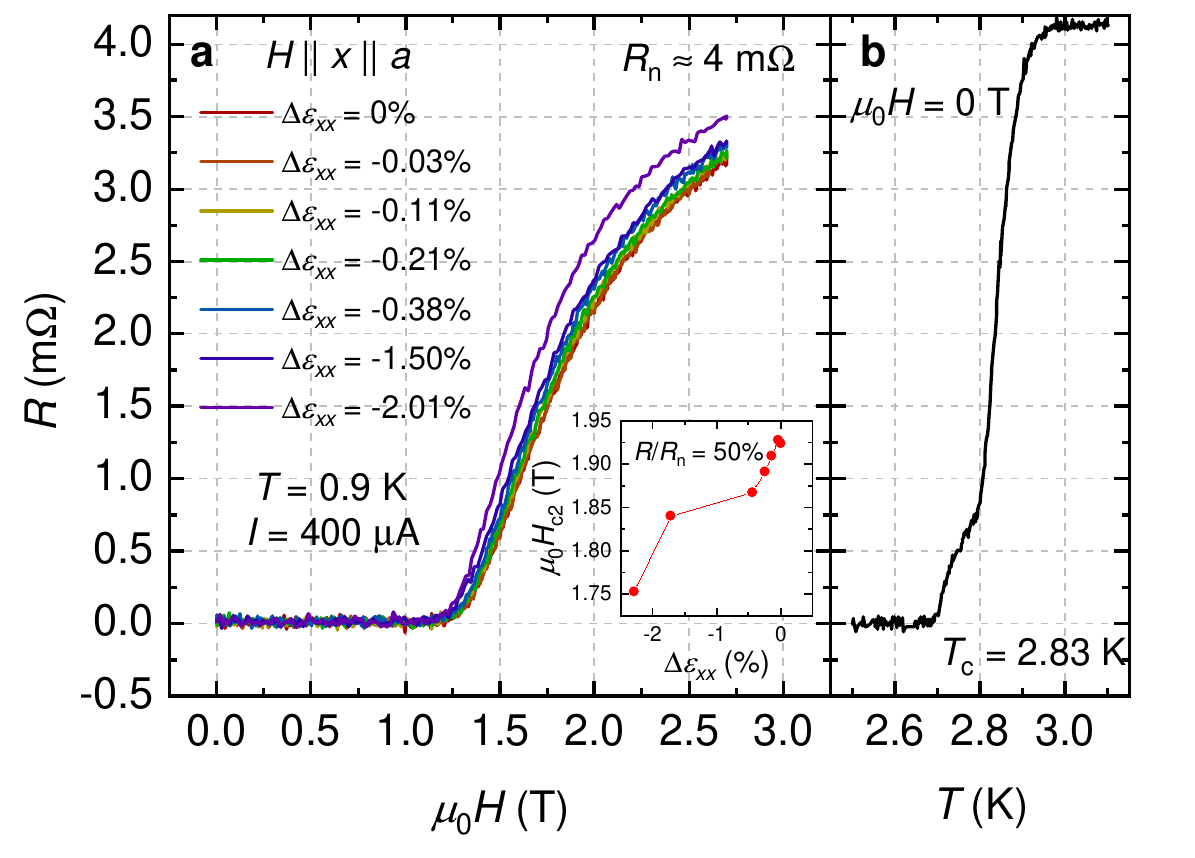}
\end{center}
\caption{
{\bf Reproducibility of the strain control of nematic superconductivity of Sample \#2.}
{\bf a}, Magnetoresistance measured at 0.9~K and for $H\parallel x$ under various strains. The upper critical field evaluated from these curves (using the criteria $R/R\subm{n} = 50$\%) is shown in the inset as a function of $\Delta\varepsilon_{xx}$. Note that with increasing compressive strain the upper critical field tends to decrease.
{\bf b}, Zero-field temperature dependence of the resistance. For this sample, \Tc\ evaluated at 50\% of the transition is 2.83~K, which is very close to that of Sample \#1.
\label{fig:Supp_RvsH_RvsT_S2}
}
\end{figure}

\begin{figure}[htb]
\begin{center}
\includegraphics[width=16cm]{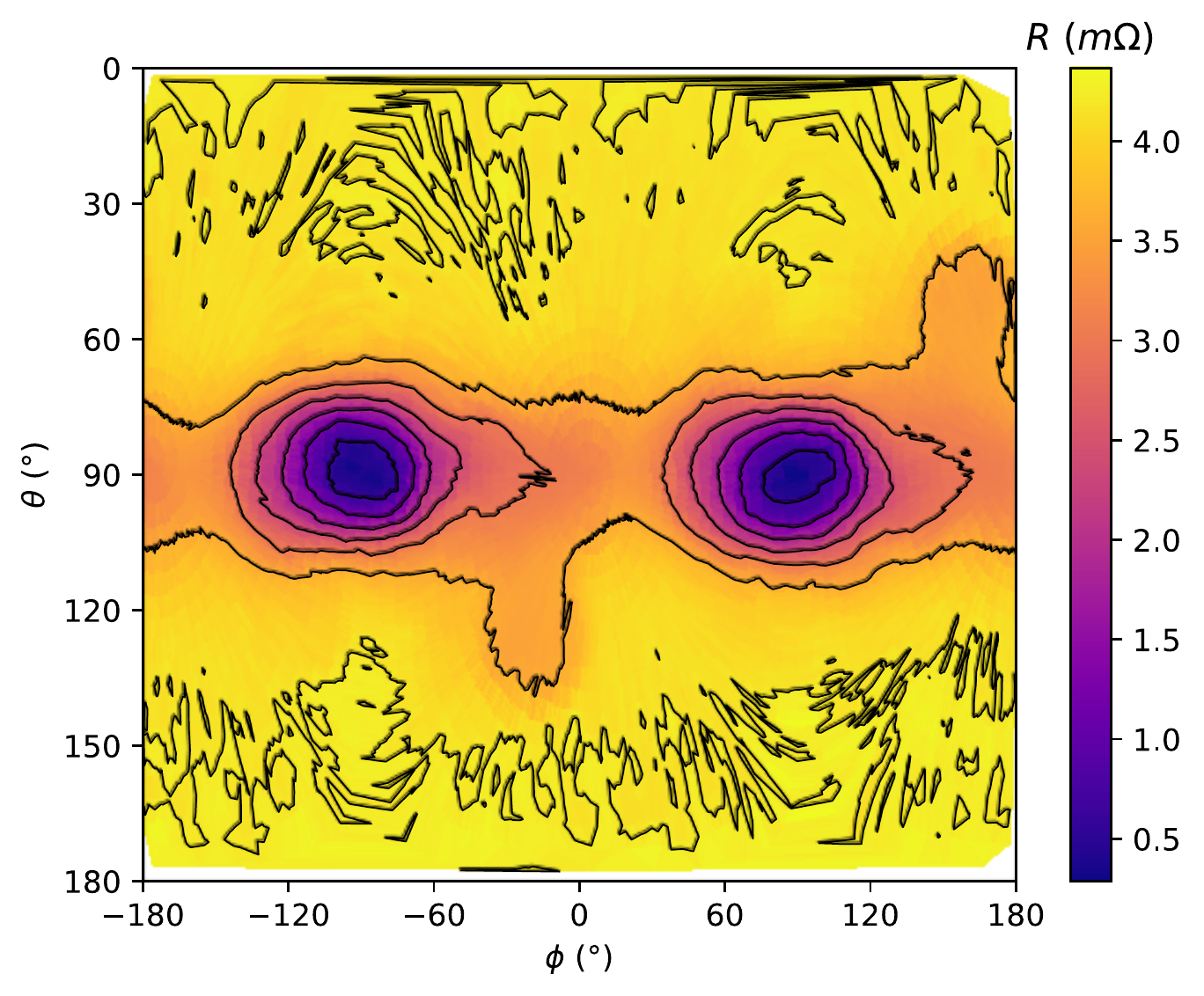}
\end{center}
\caption{
{\bf Polar and azimuthal magnetic field dependence of resistance of Sample \#2 at zero strain.}
The two-fold nematic SC component in the basal plane ($\theta=90$\deg) is clearly seen. Notice that the contours around the purple region have oval shape, nearly free from dips at $\phi_{ab} = \pm 30$ and $\pm 150$\deg. This fact indicates that contributions from minor domains are much weaker in this sample than in Sample \#1. The data here were taken at $\mu_0H=2.7$~T, ${T}=0.9$~K, and $I={}$\SI{400}{\micro\ampere}.
\label{fig:Supp_RvsPhiTheta_S2}
}
\end{figure}

It is important to show that the strain control of the nematic superconductivity is reproducibly observed in other samples. In Fig.~\ref{fig:Supp_RvsH_RvsT_S2}, we show the resistance and upper critical field (\Hcc) of another sample (now referred to as Sample \#2) of \sbsn. The strain dependence of \Hcc\ $\parallel x$ of Sample \#2 is qualitatively similar to that of Sample \#1, the sample that is mainly discussed in this Letter: a decreasing trend of \Hcc\ $\parallel x$ with compressive strain. The strain effect is less significant in this sample, likely because the sample is already in a nearly single-domain state without the external strain. Indeed, in Fig.~\ref{fig:Supp_RvsPhiTheta_S2}, the contour plot of magnetoresistance as functions of the polar and azimuthal field angles, we can see that the six-fold behavior due to minor domains is rather weak in Sample \#2, compared with the similar plot of Sample \#1 (Supplementary Fig.~S9b).
%This fact indicate that Sample \#2 is nearly in the single-domain \Y{0} state.

\clearpage

%%%%%%%%%%%%%%%%%%%%%%%%%%%%%%%%%%%%%%%%%%%%%%%%%%%%%%%%
\subsection{Strain dependence of \Tc}
\label{sec:Supp_TcvsDexx}

\begin{figure}[htb]
\begin{center}
\includegraphics[width=14cm]{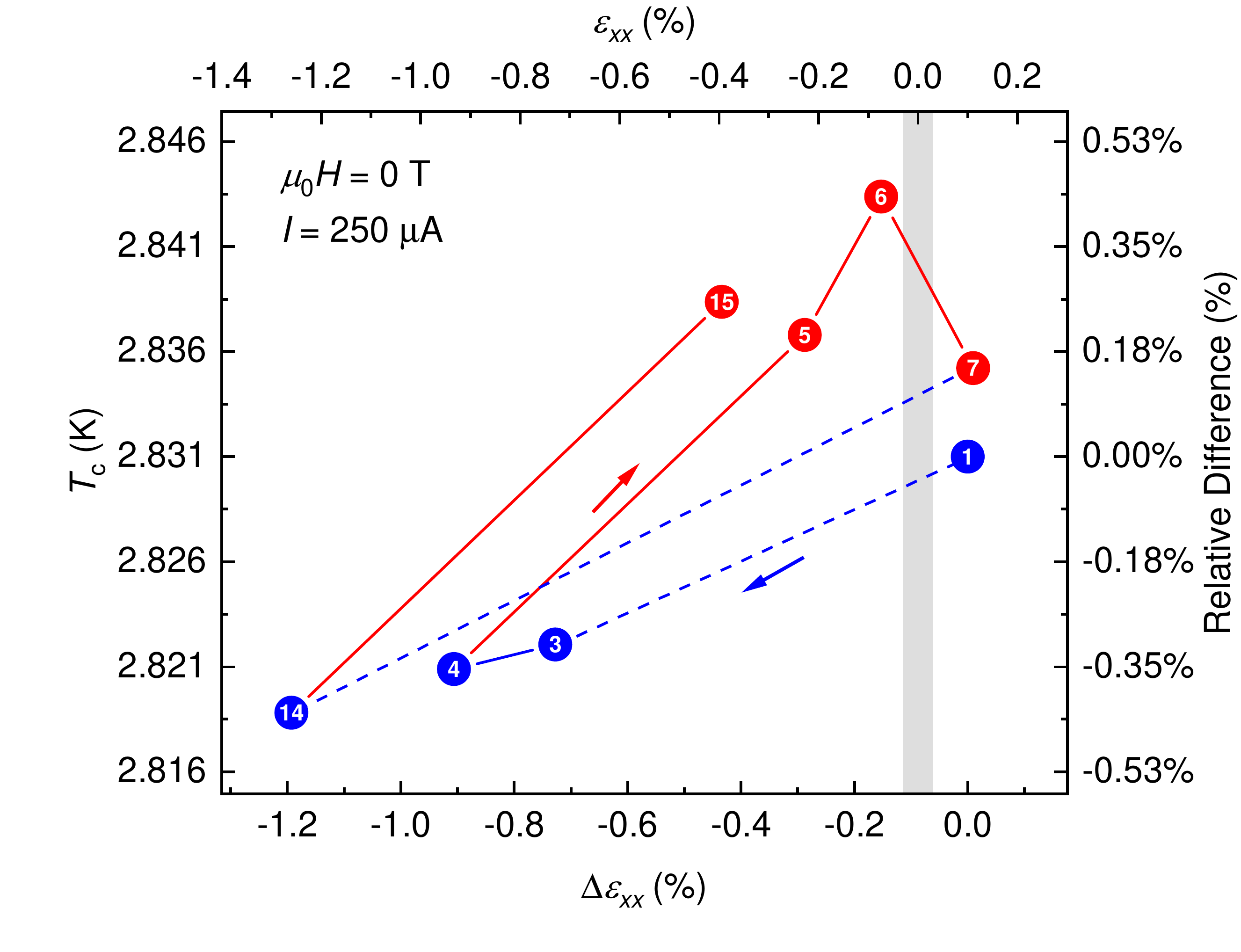}
\end{center}
\caption{
{\bf Superconducting critical temperature versus applied strain at zero field.}
The numbers in the data points indicate the order of the measurements. The lines also indicate the measurement order. Among them, the dotted lines indicate that the measurement sequence number increases by more than 1, because zero-field resistance measurement was not performed between these measurement sequences. The blue and red data points indicate the cases that the measurement was performed after a decrease and increase in applied strain, respectively. The relative difference in the right vertical axis is defined as the change in \Tc\ from the zero strain \Tc\ (2.831 K).
\label{fig:Supp_TcvsDexx}
}
\end{figure}

To see the strain dependence of superconducting properties other than the upper critical field, we show in Fig.~\ref{fig:Supp_TcvsDexx} the dependence of the superconducting critical temperature \Tc\ on the applied strain at zero field. Here, \Tc\ is defined as the midpoint of the transition. We find a decreasing trend of \Tc\ with compressive strain but the overall change is less than 1\%. 
%Interestingly, \Tc\ shows the same hysteretic trend as \Hcc\ (see Main Fig.~\ref{fig3}): upon releasing compressive strain, \Tc\ is overall larger even at zero strain. 
The change in \Tc\ may be due to a change in the density of states, as reported in the hydrostatic-pressure study of \sbsn~\cite{Nikitin2016.PhysRevB.94.144516}.

\clearpage

%%%%%%%%%%%%%%%%%%%%%%%%%%%%%%%%%%%%%%%%%%%%%%%%%%%%%%%%
\subsection{Rotation of the nematic direction by applied uniaxial strain based on the Ginzburg-Landau theory}
\label{sec:Supp_NematicSwitching}

The Ginzburg-Landau (GL) free energy on the coupling between the nematic superconductivity and a uniaxial strain under existence of a pre-existing symmetry-breaking field (SBF) is given by~\cite{How2019}
\begin{gather}
f\subm{SB} = (g\vec{\varepsilon}+g_0\vec{\varepsilon_0})\cdot\vec{S},\label{eq:fsb1}
\end{gather}
where $\vec{\varepsilon}$ is the applied strain vector, $\vec{\varepsilon_0}$ is the pre-existing SBF vector, $\vec{S}$ is the director of the nematic superconductivity, and $g$ and $g_0$ are the coupling constants.
The strain and SBF vectors are expressed as
\begin{alignat}{1}
\vec{\varepsilon} = \left(
\begin{matrix}
\varepsilon_{xx} - \varepsilon_{yy} \\
-2\varepsilon_{xy} \\
\end{matrix}
\right) = U\left(
\begin{matrix}
\cos(-2\phi) \\
\sin(-2\phi) \\
\end{matrix}
\right),\label{eq:strain}
\end{alignat}
and
\begin{alignat}{1}
\vec{\varepsilon_0} = 
U_0\left(
\begin{matrix}
\cos(-2\phi_0) \\
\sin(-2\phi_0) \\
\end{matrix}
\right),\label{eq:pre-existing-SBF}
\end{alignat}
where $U$ is the magnitude of applied anisotropic strain, $U_0$ is the magnitude of of the pre-existing SBF, $\phi$ is the angle of the strain within the basal plane of \BS, and $\phi_0$ is the angle of the pre-existing SBF.
The nematic SC order parameter is expressed as
\begin{alignat}{1}
\vec{\eta} = \left(
\begin{matrix}
\eta_x \\
\eta_y \\
\end{matrix}
\right) = \eta\left(
\begin{matrix}
\cos\phi_\eta \\
\sin\phi_\eta \\
\end{matrix}
\right).\label{eq:eta}
\end{alignat}
With this notation, the $\vec{S}$ vector has the form
\begin{alignat}{1}
\vec{S} = \left(
\begin{matrix}
|\eta_x|^2-|\eta_y|^2 \\
-2\eta_x \eta_y \\
\end{matrix}
\right)
=\eta^2\left(
\begin{matrix}
\cos(-2\phi_\eta) \\
\sin(-2\phi_\eta) \\
\end{matrix}
\right).\label{eq:S}
\end{alignat}
Substituting \eqref{eq:S}, \eqref{eq:strain}, and \eqref{eq:pre-existing-SBF} into \eqref{eq:fsb1}, then \eqref{eq:fsb1} simplifies to
\begin{gather}
f\subm{SB} = \eta^2[gU\cos(2(\phi_\eta-\phi)) + g_0U_0\cos(2(\phi_\eta-\phi_0))].
\end{gather}
The nematicity direction $\phi_\eta$ is chosen such that the free energy is minimized (i.e. $d f\subm{SB}/{d \phi_\eta}=0$ and ${d^2 f\subm{SB}}/{d \phi_\eta^2}>0$):
% \begin{gather}
% \phi_\eta = \phi - \frac{1}{2}\arctan\left(\frac{\sin(2(\phi-\phi_0))}{gU/g_0U_0 + \cos(2(\phi-\phi_0))}\right) + \frac{\pi}{2}k, \label{eq:phi_eta}
% \end{gather}
\begin{gather}
\Delta\phi_\eta = -\frac{1}{2}\arctan\left(\frac{\sin(2(\Delta\phi_0))}{gU/g_0U_0 + \cos(2(\Delta\phi_0))}\right) + \frac{\pi}{2}k, \label{eq:phi_eta}
\end{gather}
where $\Delta\phi_\eta \equiv \phi_\eta - \phi$ is the nematicity direction with respect to the applied strain direction, $\Delta\phi_0 \equiv \phi - \phi_0$ is the angle between the strain and the pre-existing SBF, and $k$ is an integer chosen such that ${d^2 f\subm{SB}}/{d \phi_\eta^2}>0$ is satisfied. 

\begin{figure}[htb]
\begin{center}
\includegraphics[width=12cm]{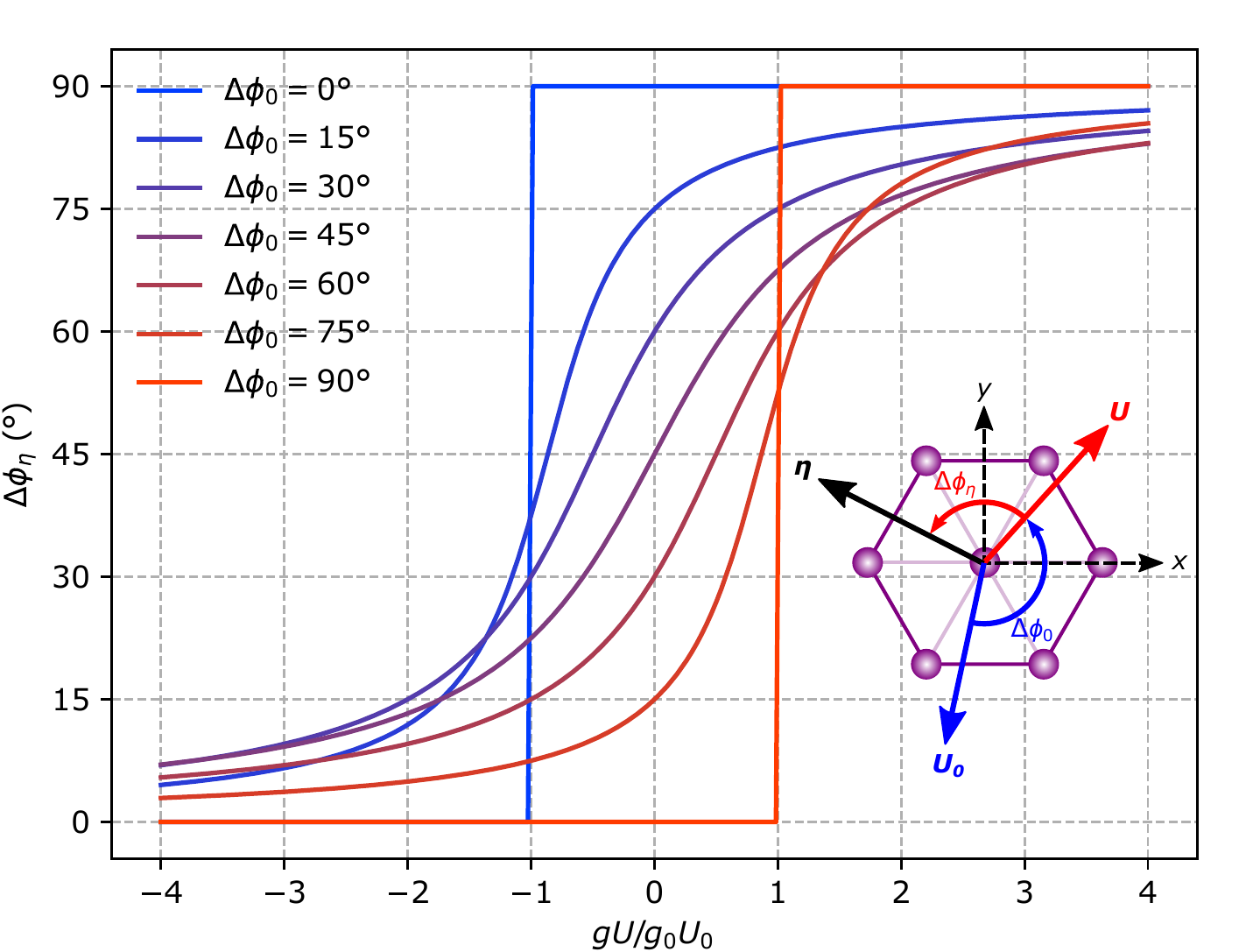}
\end{center}
\caption{
{\bf Rotation of the nematic direction by applied uniaxial strain.}
The uniaxial strain with the magnitude $U$ is applied at an angle $\Delta\phi_0$ from the pre-existing symmetry breaking field (strength $U_0$). $\Delta\phi_\eta$ is the angle relative to the axis of the applied strain.
The inset shows the definitions of the axes and the symmetry breaking fields ($U$, $U_0$) and nematic director ($\eta$) with respect to the crystal structure in the $ab$ plane.
%The uniaxial strain with the magnitude $U$ is applied along the \textit{x} axis ($\phi=0$\deg) whereas the pre-existing symmetry breaking field (strength $U_0$) is assumed to be along the other nearest \textit{a} axis ($\phi_0=60$\deg). The color indicates the nematic order: \Dx\ (blue; 0\deg; $\parallel a$) and \Dy\ (green; 30\deg; $\parallel a^*$). Intermediate angle values indicates that the nematicity is neither aligned along the $a$ or $a^*$ axes.
\label{fig:Supp_NematicSwitching}
}
\end{figure}

The result of the above equation \eqref{eq:phi_eta} is shown in Fig.~\ref{fig:Supp_NematicSwitching}. It is evident that, when the angle between the applied strain and the pre-existing SBF is orthogonal ($\Delta\phi_0=90$\deg), the nematic direction changes discontinuously when the applied strain term is equal to the pre-existing SBF term (i.e. $gU=g_0U_0$). When the pre-existing SBF is parallel ($\Delta\phi_0=0$\deg), the nematic direction changes suddenly when the applied strain term is equal and opposite to the pre-existing SBF term (i.e. $gU=-g_0U_0$). At intermediate angles, the direction of the nematicity changes continuously but rapidly until the ratio of the applied and pre-existing SBF, $gU/g_0U_0$, reaches about 1, after which it changes more gradually.
This rotation is very likely the driving force of the nematic domain change under uniaxial compression observed in this work.

\clearpage

%%%%%%%%%%%%%%%%%%%%%%%%%%%%%%%%%%%%%%%%%%%%%%%%%%%%%%%%
\subsection{Transformation between the laboratory and sample frames}
\label{sec:Supp_EulerTransformExample}

In this section, we describe the procedure to determine the transformation relation between the sample and laboratory frames, to align magnetic fields accurately with respect to the crystalline axes. 

\begin{figure}[htb]
\begin{center}
\includegraphics[width=11cm]{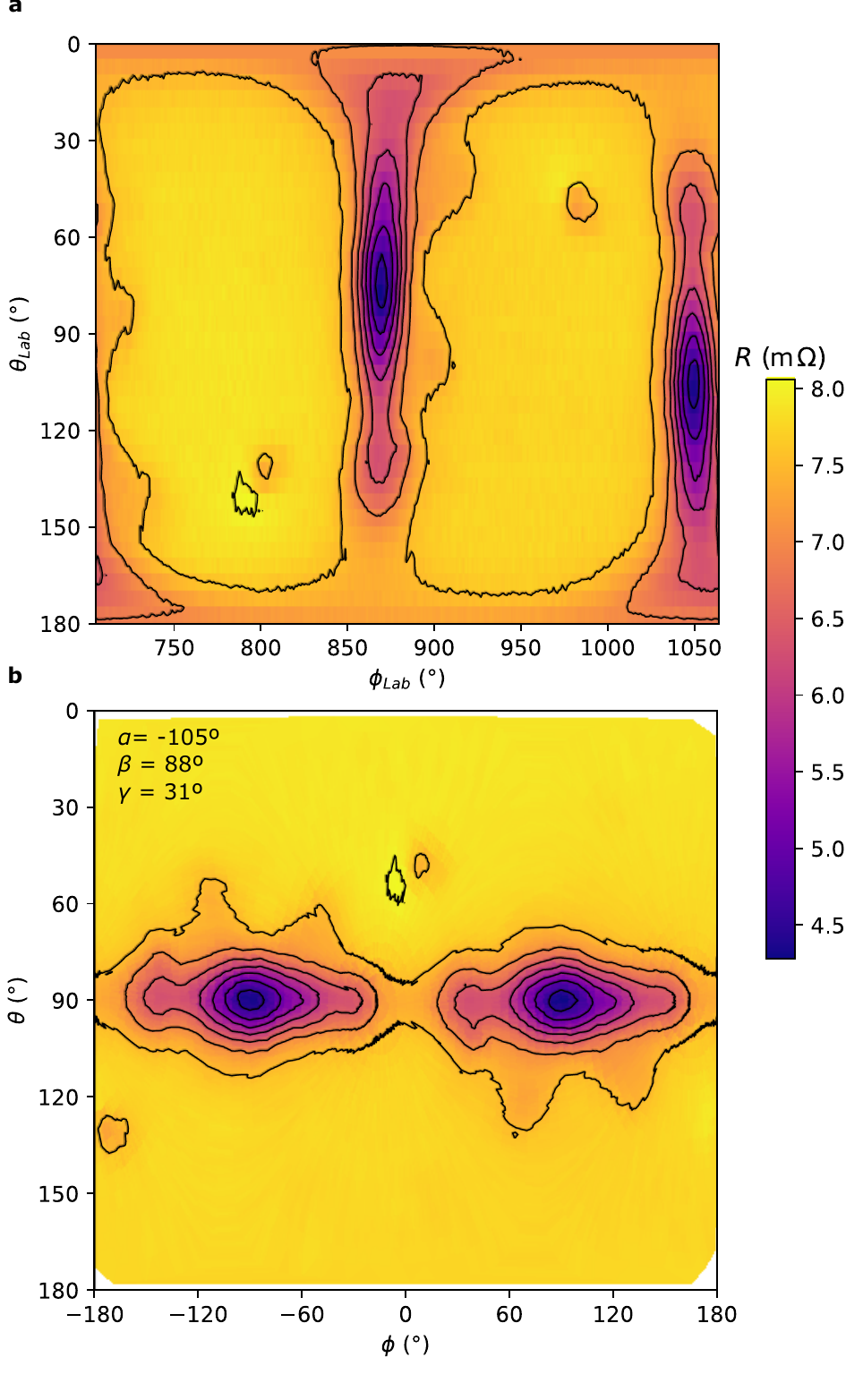}
\end{center}
\caption{
{\bf Euler transform from laboratory to sample frame.}
{\bf a}, Contour plot of the angular magnetoresistance plotted as functions of the azimuthal and polar field angles in the laboratory frame. The data were obtained at 2.2~K and 1~T.
{\bf b}, Same data plotted as functions of field angles in the sample frame, after rotating the data by using the Euler angles ($\alpha=-105$\deg, $\beta=88$\deg, $\gamma=31$\deg).
\label{fig:Supp_EulerTransformExample}
}
\end{figure}

In this work, the magnetic field was applied using a vector-magnet system, which consists of two orthogonal superconducting magnets: one pointing in the vertical direction and the other in the horizontal direction in the laboratory frame~\cite{Deguchi2004}. The polar and azimuthal angles of the magnetic field are indicated by $\theta\subm{Lab}$ and $\phi\subm{Lab}$, respectively. To know the transformation between the laboratory frame angles ($\theta\subm{Lab}$ and $\phi\subm{Lab}$) and the sample frame angles ($\theta$ and $\phi$), we made use of the anisotropy in \Hcc. We first measured the angular magnetoresistance in the superconducting transition region, covering the full $4\pi$ solid angle of the magnetic field, as shown in Fig.~\ref{fig:Supp_EulerTransformExample}a in the laboratory frame. Because \Hcc\ of \sbs\ is smallest along the $c$ axis~\cite{Pan2016.SciRep.6.28632}, the field direction with the largest resistance is the $c$ direction and the plane with relatively small resistance should be the $ab$ plane. If this data is correctly transformed into the sample frame by using a $3 \times 3$ rotation matrix $R$, the former should be located at $\theta = 0$ or 180\deg, and the latter should lie at $\theta = 90$\deg. 
Thus, our goal is to find such a matrix $R$.

In general, a vector in the laboratory frame $v\subm{Lab}$ transforms to a vector in the sample frame $v$ via:
\begin{gather}
v = R \cdot v\subm{Lab}.
\end{gather}
The vectors $v$ and $v_{Lab}$ are in Cartesian coordinates. The rotation matrix $R$ can be decomposed into three elemental rotation matrices with the Euler angles $\alpha$, $\beta$, and $\gamma$ :
\begin{gather}
R = Z(\gamma)X(\beta)Z(\alpha),
\end{gather}
which corresponds to the combination of a rotation by $\alpha$ about $z$ axis, then a rotation by $\beta$ about the rotated $x$ axis, and then a rotation by $\gamma$ about the rotated $z$ axis.
Note that the elemental rotation matrices are given as follows:
\begin{alignat}{1}
X(\omega) &= \begin{pmatrix}
1 & 0 & 0 \\
0 & \cos \omega &  -\sin \omega \\[3pt]
0 & \sin \omega  &  \cos \omega \\[3pt]
\end{pmatrix} \\[6pt]
Z(\omega) &= \begin{pmatrix}
\cos \omega &  -\sin \omega & 0 \\[3pt]
\sin \omega &   \cos \omega & 0\\[3pt]
0 & 0 & 1\\
\end{pmatrix},
\end{alignat}
where $\omega$ is one of the Euler angles. Lastly, the sample frame vector $v$ is converted from Cartesian to spherical coordinates defined by the two variables $\theta$ and $\phi$. The basal plane $\phi_{ab}$ is given as $\phi$ on the plane of $\theta=90$\deg. 

When determining the Euler angles from the experiment, we first find the plane of low resistance (i.e. the $ab$ plane) comes on the plane $\theta =90$\deg\ when we used the Euler angles $\alpha=-105$\deg\ and $\beta=88$\deg. To determine $\gamma$, we need to use the fact that the sample's $x$ axis (one of the $a$ axes) is roughly oriented along $\theta_{Lab}=0$\deg, as described in Methods. 
This $x$ axis should be transformed to $\phi=0$\deg\ in the $\theta=90$\deg\ plane after the $\gamma$ rotation. 
This determines the last Euler angle $\gamma$ to be 31\deg.
With this combination of the Euler angles, the angular magnetoresistance is now transformed as shown in Fig.~\ref{fig:Supp_EulerTransformExample}b, matching with the expectation explained above.
%we first observe that the sample's $a$ axis is roughly oriented along $\theta_{Lab}=0$\deg, as seen in Fig.~\ref{fig:Supp_EulerTransformExample}. The laboratory frame is first rotated so that the $a$ axis is roughly at $\phi=0$\deg\ and $\theta=90$\deg, then it is rotated so that the two peaks are at $\theta=90$\deg\ and $\phi=\pm90$\deg. For this experiment the Euler angles are $\alpha=-105$\deg, $\beta=88$\deg, $\gamma=31$\deg.

Finally, the magnetic field coordinate matrices expressed in the sample frame $H\subm{Sample}$ and that in the laboratory frame $H\subm{Lab}$ can be converted back and forth via the relation
\begin{gather}
H\subm{Sample} = R \cdot H\subm{Lab}.
\end{gather}
%where the coordinate matrices have the following form
%\begin{alignat}{1}
%H = \left(
%\begin{matrix}
%x_1, x_2, \ldots, x_N \\
%y_1, y_2, \ldots, y_N \\
%z_1, z_2, \ldots, z_N \\
%\end{matrix}
%\right).
%\end{alignat}
  
As mentioned in Methods, the magnetic field presented in the Main Text are all expressed in the Sample frame determined in this way.

\clearpage

%%%%%%%%%%%%%%%%%%%%%%%%%%%%%%%%%%%%%%%%%%%%%%%%%%%%%%%%
\subsection{Angular magnetoresistance at zero applied strain}
\label{sec:Supp_R_Phi_Theta}

In this section, we show angular magnetoresistance covering the whole 4$\pi$ solid angles of the field directions at zero applied strain, in order to demonstrate that the observed behavior is not due to the field misalignment.

\begin{figure}[htb]
\begin{center}
\includegraphics[width=16cm]{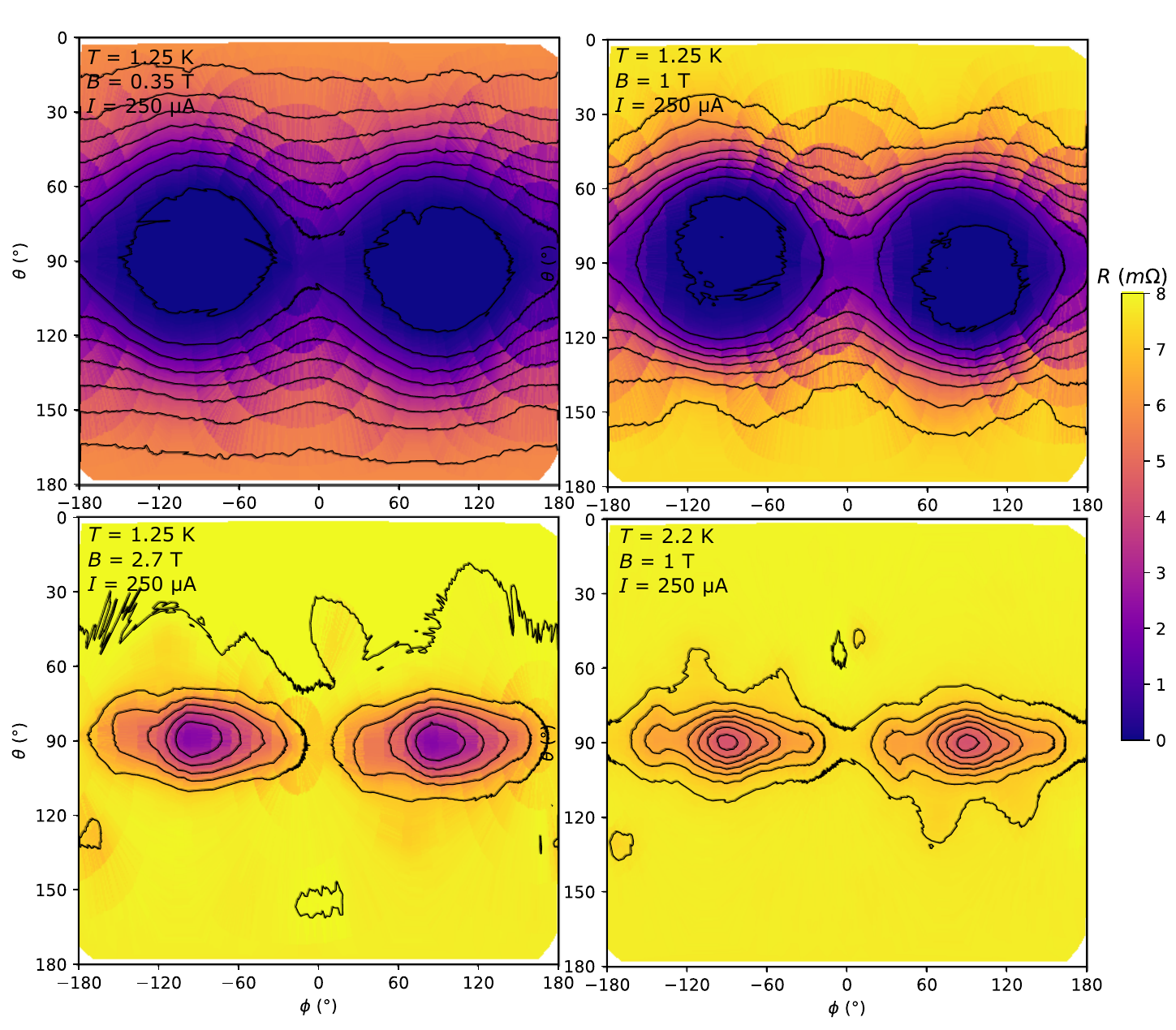}
\end{center}
\caption{
{\bf Polar and azimuthal angle dependences of the magnetoresistance at zero strain for various magnetic-field and temperature conditions.}
The light yellow and dark blue regions correspond to normal state and superconducting state, respectively. Measurement conditions are indicated in the top-left corner of each panel.
%Note that at high field and/or high temperature two prominent peaks remain, indicating superconducting regions with large upper critical field.
\label{fig:Supp_R_Phi_Theta}
}
\end{figure}

In Fig.~\ref{fig:Supp_R_Phi_Theta}, we show the colour plots of the magnetoresistance as functions of the polar and azimuthal field angles, measured at different magnetic field strength and temperature at zero applied strain (i.e. $\Delta\varepsilon_{xx} =0$\%). Evidently, for all cases the strong two-fold behavior along the $\phi$ direction due to the nematic superconductivity is seen. For low temperature and/or low field, most of the angles are largely in the superconducting state (corresponding to the dark-blue region), whereas for high temperature or high field only the regions with largest upper critical field remain in the superconducting state. From these data, we confirm that our alignment of the magnetic field to the crystal axis is quite accurate and field-misalignment effect is negligible.

In addition to the strong two-fold behavior, the data near the onset (the two bottom panels of Fig.~\ref{fig:Supp_R_Phi_Theta}) exhibit small anomalies at around $\phi = \pm 30$\deg\ and $\pm 150$\deg. See that some contours have dips at these angles. These anomalies are due to the existence of nematic subdomains as discussed in the Main Text.

\clearpage

%%%%%%%%%%%%%%%%%%%%%%%%%%%%%%%%%%%%%%%%%%%%%%%%%%%%%%%%
\subsection{Model simulation}
\label{sec:Supp_3DCircuit}

In order to simulate the magnetoresistance and the upper-critical-field behavior under single and multiple nematic SC domains, we performed a model simulation. In this section, details of the simulation will be discussed.

\subsubsection*{Magnetoresistance of each domain}

\begin{figure}[htb]
\begin{center}
\includegraphics[width=11cm]{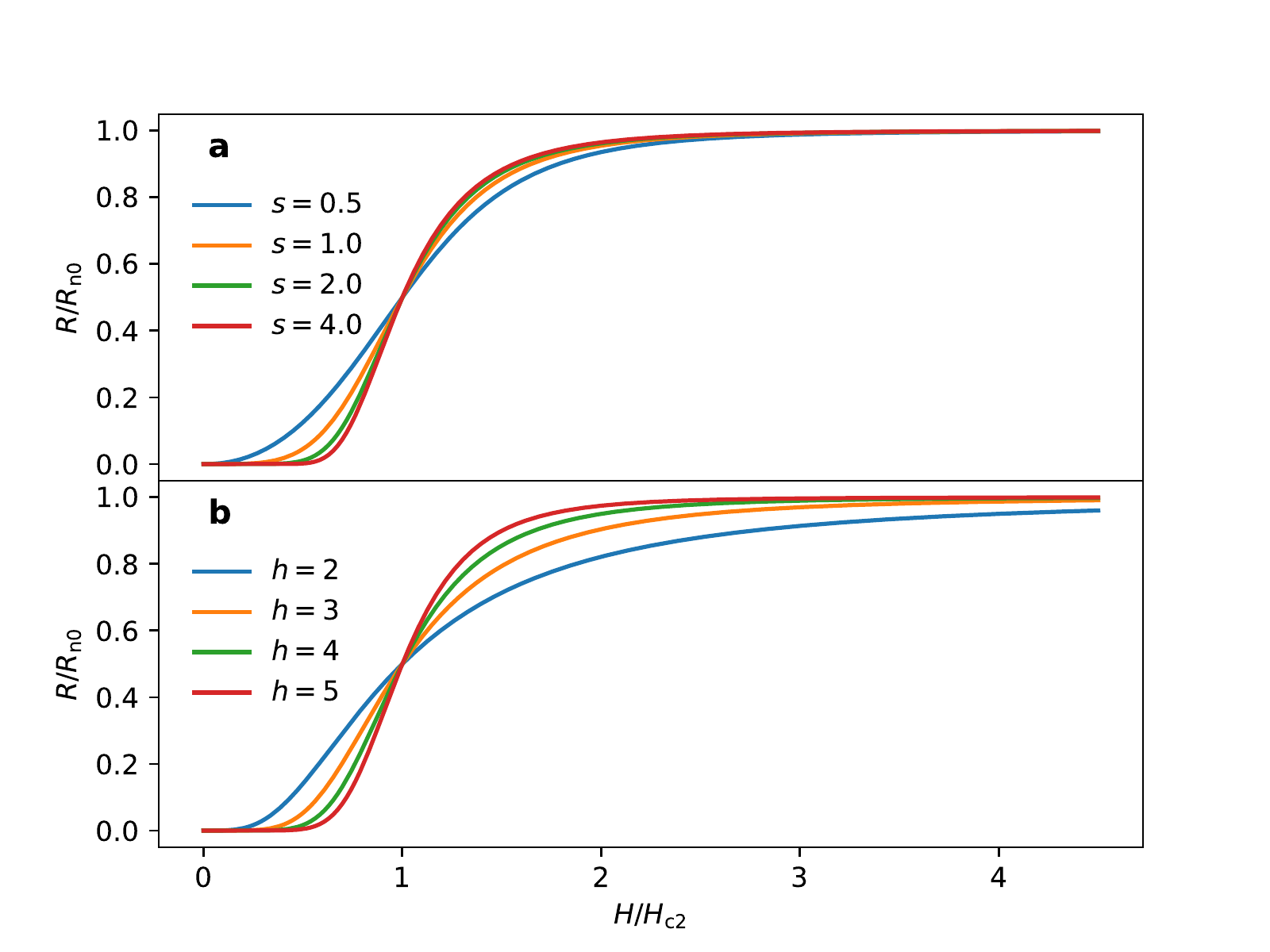}
\end{center}
\caption{
{\bf Plot of Eq.~\eqref{eq:1}, representing resistance versus applied magnetic field for a single domain, for varying parameters.}
{\bf a}, For varying $s$ where $h=4.39$.
{\bf b}, For varying $h$ where $s=2.04$.
%For calculation we assume that $\phi_{ab}=0$\deg\ and $H\subm{c2}(\phi_{ab}=0)=\frac{1}{3}$.
\label{fig:Supp_RRn_v_B_Sim_h_s}
}
\end{figure}

Firstly, we have to define the magnetoresistance behavior of each domain. We assumed that the magnetoresistance of a single nematic SC domain obeys the following empirical equation:
\begin{gather}
%R(H)/R\subm{n0} =(1+(H/H_0 )^{-h(H_{c2})})^{-s},\label{eq:1}
\frac{R(H, \phi_{ab})}{R\subm{n0}} =\left[ 1+\left(\frac{H}{H\subm{c2}(\phi_{ab})} \right)^{-h(H\subm{c2}(\phi_{ab}))} (2^{1/s} - 1) \right] ^{-s},\label{eq:1}
\end{gather}
where $R\subm{n0}$ is the normal state resistance of the domain,  $\phi_{ab}$ is the in-plane field angle, 
\Hcc\ is the upper critical field (midpoint), $s$ is an exponent determining the shape of the $R(H)$ curve around \Hcc, and $h(H\subm{c2})$ is another exponent introduced to depict the \Hcc-dependent transition width. Notice that the coefficient $2^{1/s} -1 $ is a correction factor to make the right-hand side of eq.~\eqref{eq:1} to 1/2 at $H = H\subm{c2}(\phi_{ab})$. 
The functional form of the right-hand side of eq.~\eqref{eq:1} is shown in Fig.~\ref{fig:Supp_RRn_v_B_Sim_h_s}.

For the actual simulation, we used the exponent $s=2.03535$ and we employed an empirical relation $h(H\subm{c2})=-2.696 \times H\subm{c2} (\phi_{ab})/H\subm{c2,max} + 5.74248$ to reproduce the observed resistance behavior of the actual sample, in particular the $H\subm{c2}$-dependent broadening of the transition. Here, $H\subm{c2,max}$ is the maximum $H\subm{c2}$ within the $ab$ plane. The angular dependence of $H_{c2}$ is approximated by the anisotropic mass model:
\begin{gather}
H\subm{c2}(\phi_{ab}) = \frac{H\subm{c2,max}}{\sqrt{\cos^2(\phi_{ab}-\phi_0)+\varGamma^2\sin^2(\phi_{ab}-\phi_0)}},
\end{gather}
where $\varGamma \equiv H\subm{c2,max}/H\subm{c2,min}$ is the anisotropy, and $H\subm{c2,min}$ is the minimum $H_{c2}$ given by $\phi_{ab}=\phi_{0}\pm\pi/2$.
The value of $\phi_{0}$ defines the nematic superconducting domains: For the $Y_n$ domain ($n = 0, 1, 2$), $\phi_{0}$ is given by $\pi/2 + n\pi/3$.
We found that $\varGamma = 3$ best reproduces the experimental data.
Thus, this $\varGamma$ value is used hereafter.
Another important parameter, $H\subm{c2,max}$ is set to 1~T unless explicitly mentioned, to reproduce the $R(H)$ curve at 2.2~K.

With these formulations and parameters, eq.~\eqref{eq:1} exhibits functional forms shown in  Fig.~\ref{fig:Supp_RRnvsB} in the case of $\phi_0 = \pi/2$ (\Y{0} domain).  Comparing these curves with the raw data shown in Fig.~1c, we can see that eq.~\eqref{eq:1} well reproduces the observed magnetoresistance of the strained sample (corresponding to a single \Y{0} domain state). Thus, the formulation described above should be valid for the simulation.

\subsubsection*{Circuit model of multiple domains}

Next, we have to assume a certain circuit to model the distribution of domains.
In Fig.~\ref{fig:Supp_3DCircuit}, we present the electrical circuit model used to produce data in Fig.~4, consisting of a 3D network of twelve resistive elements $R\subm{1a}$, $R\subm{1b}$, $\ldots$, $R\subm{2B}$. The end-to-end total resistance of this circuit $R\subm{total}$ is given by a certain function $f$:
\begin{gather}
R\subm{total}(H, \phi_{ab}) = f\left(  R\subm{1a}(H, \phi_{ab}),R\subm{1b}(H, \phi_{ab}),...,R\subm{2B}(H, \phi_{ab})  \right),\label{eq:Rtotal}
\end{gather}
which is determined by standard techniques of circuit analysis. 
The normal-state resistance $R\subm{total,n}$ of the net circuit is given by 
\begin{gather}
    R\subm{total,n} = R\subm{total}(H\to \infty)
\end{gather}

\begin{figure}[htb]
\begin{center}
\includegraphics[width=16cm]{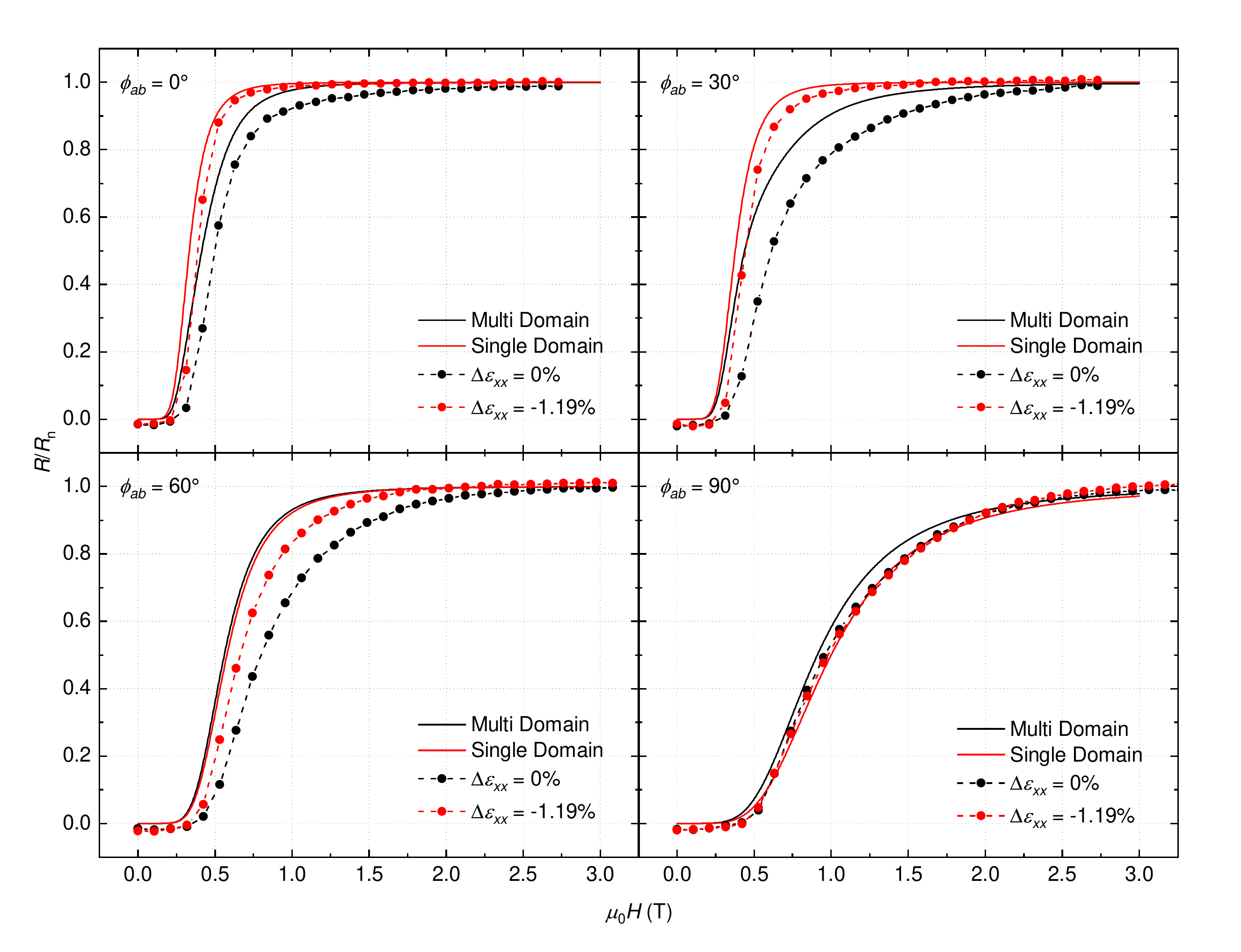}
\end{center}
\caption{
{\bf Comparison between data and model for $R/R_n$ vs $\mu_0H$ at different in-plane angles.}
The angle of the magnetic field is shown by $\phi_{ab}$ in the top-left corner in each panel.
The multi domain model is given by eq.~\eqref{eq:Rtotal}.
The single domain model is given by eq.~\eqref{eq:1}.
The nematic anisotropy used is $\varGamma = 3$.
The maximum in-plane upper critical field is $\mu_0H\subm{c2,max}=1$~T.
\label{fig:Supp_RRnvsB}
}
\end{figure}

\begin{figure}[htb]
\begin{center}
\includegraphics[width=10cm]{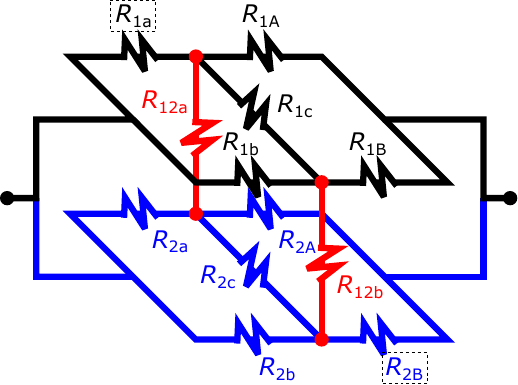}
\end{center}
\caption{
{\bf Electrical circuit of a 3D network of resistive elements used for the simulation.}
The black and blue colored lines represent the top and bottom parts of the circuit, respectively. The red lines are the interconnecting paths between the top and bottom parts of the circuit. 
%The dots terminating on the leftmost and rightmost side of the circuit are the terminals used for determining the equivalent resistance. 
The elements marked by the dotted box are set to the minor domains in case of the multi-domain simulation.
\label{fig:Supp_3DCircuit}
}
\end{figure}

For the multi-domain simulation presented in Fig.~4 of the Main Text, we assumed that $R\subm{1a}$ and $R\subm{2B}$ are $Y_1$ and $Y_2$ domains and the rest are $Y_0$ domains. For each domain, the normal-state resistance value $R\subm{n0}$ is assumed to be the same. Magnetoresistance curves obtained for this multi-domain case are shown in Fig.~\ref{fig:Supp_RRnvsB}, which captures features of the magnetoresistance of the unstrained sample (i.e. $\Delta\varepsilon_{xx} = 0$).
For the single-domain simulation, we set all components to the \Y{0} domain. We comment that, for the single domain case, $R\subm{total}/R\subm{total,n}$ is identical to $R/R\subm{n0}$ of eq.~\eqref{eq:1}. 

\subsubsection*{Illustrative explanation using a simpler model}

\begin{figure}[htb]
\begin{center}
\includegraphics[width=10cm]{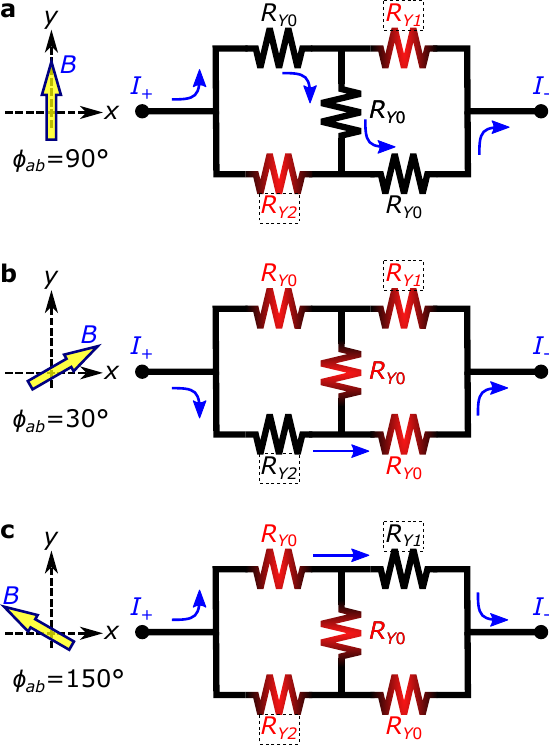}
\end{center}
\caption{
{\bf Simplified electrical circuit diagrams describing the current path dependence on the direction of the applied field.}
The black and red color of the resistive element (R) indicate relatively low and high resistance, respectively. The blue arrows indicate the path of the current at each configuration. The yellow arrow indicates the angle ($\phi_{ab}$) of the magnetic field ($B$) relative to the sample's \textit{x} axis. The three domains are $Y_0$, $Y_1$, and $Y_2$, which are most superconducting at the field angles $\pm90$\deg, $150$\deg ($-30$\deg), and $30$\deg ($-150$\deg).
{\bf a}, At $\phi_{ab}=90$\deg\ $R_{Y0}$ is least resistive and the current passes through the center avoiding $R_{Y1}$ and $R_{Y2}$.
{\bf b}, At $\phi_{ab} =30$\deg\ $R_{Y2}$ is least resistive so the current initially avoids $R_{Y0}$ but after the first element the current has to pass through the less resistive path that is $R_{Y0}$ (instead of $R_{Y0}$ and $R_{Y1}$).
{\bf c}, same as (b) except for $\phi_{ab} =150$\deg\ and $R_{Y1}$ is initially preferred. The dotted box indicates the minor domain.
\label{fig:Supp_2DCircuit}
}
\end{figure}

In order to illustrate how the path of the current changes depending on the direction of the applied magnetic field, we show a simplified version of the above circuit in Fig.~\ref{fig:Supp_2DCircuit}. When the direction of the applied magnetic field is parallel to the axis that has the largest \Hcc\ for the dominant domain, $Y_0$ (Fig.~\ref{fig:Supp_2DCircuit}a), then the $Y_0$ domain has lower resistance than the minor domains $Y_1$ and $Y_2$ and hence the current passes mostly through the $Y_0$ domains. If the field angle is aligned with the \Hcc\ maximum of either the $Y_1$ or $Y_2$ domains (Fig.~\ref{fig:Supp_2DCircuit}b or c), then the current will certainly pass through those domains. However, due to the configuration of the domains in the network, the current must pass through a $Y_0$ domain as well to reach the opposite end. This effect is what ensures that the minor domains have a relatively smaller influence on \Hcc\ than the dominant domain except for the very vicinity of the onset of superconductivity. Hence we get the characteristic 6-fold in-plane \Hcc\ with one of the 2-fold \Hcc\ being relatively larger than the other near the onset (95 or 80\% criteria \Hcc) but purely two-fold behavior close to zero resistance state (20 or 5\% criteria \Hcc). 

\clearpage

%%%%%%%%%%%%%%%%%%%%%%%%%%%%%%%%%%%%%%%%%%%%%%%%%%%%%%%%
\subsection{\textit{H-T} phase diagrams}
\label{sec:Supp_Hc2vsT_AnivsT}

\begin{figure}[htb]
\begin{center}
\includegraphics[width=15cm]{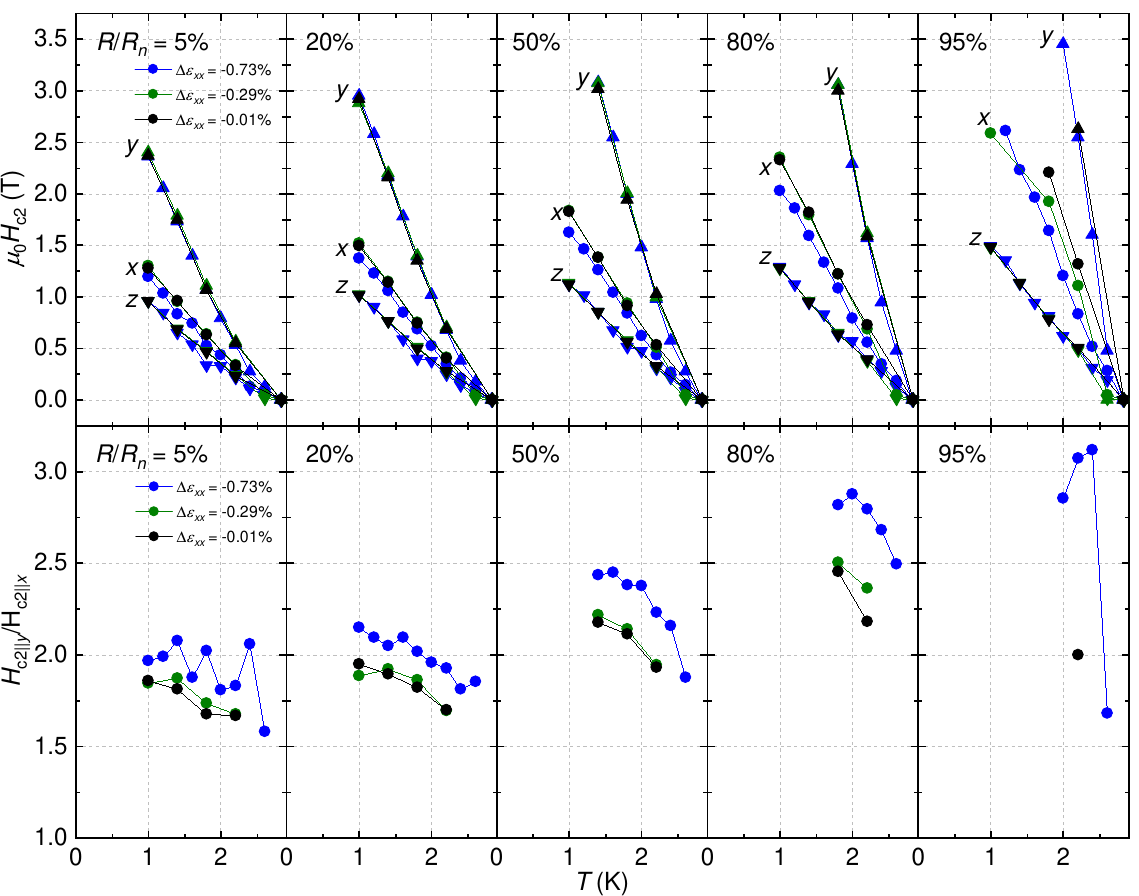}
\end{center}
\caption{
{\bf Upper critical field \Hcc\ (top panels) and in-plane \Hcc\ anisotropy (bottom panels) dependence on the temperature for varying applied strain.}
The colors black, green, and blue correspond to small to larger compressive strains. The numbers in the top corner of each sub-panel indicates the criteria used for determining \Hcc. Note that at low temperatures there are missing data points due to the magnetoresistance data not having the necessary resistive range for the specified \Hcc\ criterion.
\label{fig:Supp_Hc2vsT_AnivsT}
}
\end{figure}

To describe the temperature evolution of the strain effect, we show 
in Fig.~\ref{fig:Supp_Hc2vsT_AnivsT} the temperature dependence of \Hcc\ determined with various $R/R\subm{n}$ criteria along the three principal axes (\textit{x}, \textit{y}, \textit{z}).
We also show the in-plane \Hcc\ anisotropy ($H\subm{c2} \parallel y / H\subm{c2} \parallel x$) under various strain in the bottom panels. With lowering temperature, \Hcc\ exhibits linear increase. The in-plane \Hcc\ anisotropy ranges 2-3 depending on the \Hcc\ criteria. Such a relatively large anisotropy is consistent with previous studies on Sr-doped \BS~\cite{Pan2016.SciRep.6.28632,Du2017.SciChinaPhysMechAstron.60.037411}. With an increase of compressive strain, the \Hcc\ anisotropy increases for the whole temperature range investigated. Thus, the uniaxial-strain control of nematic superconductivity is achieved irrespective of the temperature range.

%\clearpage

%\bibliography{References}

% \end{document}

\end{document}